\documentclass[12pt]{article}
\pdfoutput=1
\usepackage{graphicx}
\usepackage{color,amssymb,amsmath}
\usepackage[a4paper, total={16.5cm,22cm}]{geometry}
\usepackage{cite}
\usepackage[colorlinks=true,linkcolor=dark_blue,
  citecolor=dark_red,urlcolor=dark_red,linktocpage=false]{hyperref}
\definecolor{dark_red}{rgb}{0.6,0,0}
\definecolor{dark_blue}{rgb}{0,0,0.6}

\usepackage{slashed}

\begin{document}

\newcommand{\lrf}[2]{ \left(\frac{#1}{#2}\right)}
\newcommand{\lrfp}[3]{ \left(\frac{#1}{#2} \right)^{#3}}
\newcommand{\vev}[1]{\langle #1\rangle}
\newcommand{\hf}{\mathchar`-}

\begin{titlepage}

\begin{center}
\hfill UT-18-1\\
\hfill  IPMU18-0036\\

\vskip 2cm

{\Large \bf 
Supersymmetric Flaxion
}

\vskip 1.5cm

{\large Yohei Ema$^{(a)}$, Daisuke Hagihara$^{(a)}$, Koichi Hamaguchi$^{(a,b)}$,\\
\vskip 0.2cm
Takeo Moroi$^{(a,b)}$ and Kazunori Nakayama$^{(a,b)}$}

\vskip 0.5cm

$^{(a)}${\em Department of Physics, 
The University of Tokyo,  Tokyo 113-0033, Japan}

\vskip 0.2cm

$^{(b)}${\em Kavli Institute for the Physics and Mathematics of the Universe (Kavli IPMU), \\
University of Tokyo, Kashiwa 277--8583, Japan
}

\end{center}

\vskip 1.5cm

\begin{abstract}
Recently, a new minimal extension of the Standard Model has been proposed, where a spontaneously broken, flavor-dependent global U(1) symmetry is introduced. It not only explains the hierarchical flavor structure in the quark and lepton sector, but also solves the strong CP problem by identifying the Nambu-Goldstone boson as the QCD axion, which we call flaxion. In this work, we consider supersymmetric extensions of the flaxion scenario. We study the CP and flavor violations due to supersymmetric particles, the effects of R-parity violations, the cosmological gravitino and axino problems, and the cosmological evolution of the scalar partner of the flaxion, sflaxion. We also propose an attractor-like inflationary model where the flaxion multiplet contains the inflaton field,
and show that a consistent cosmological scenario can be obtained, including inflation, leptogenesis, and dark matter.
\end{abstract}

\end{titlepage}

\renewcommand{\thepage}{\arabic{page}}
\setcounter{page}{1}
\renewcommand{\thefootnote}{$\natural$\arabic{footnote}}
\setcounter{footnote}{0}

\newpage

\section{Introduction} 

Recently, a new minimal extension of the Standard Model has been proposed~\cite{Ema:2016ops,Calibbi:2016hwq}, 
where the Froggatt-Nielsen U(1) flavor symmetry~\cite{Froggatt:1978nt} and the Peccei-Quinn (PQ) symmetry~\cite{Peccei:1977hh} are identified:
\begin{align}
{\rm U}(1)_{\text{F}} = {\rm U}(1)_{\text{PQ}}.
\end{align}
In this scenario, flavor dependent U(1) charges of the quarks and
leptons can explain their hierarchical masses and mixings, while the
Nambu-Goldstone (NG) boson of the spontaneously broken U(1) symmetry
plays the role of the QCD axion~\cite{Weinberg:1977ma,Wilczek:1977pj}
to solve the strong CP problem as well as the dark matter
puzzle~\cite{Preskill:1982cy,Abbott:1982af,Dine:1982ah}.  We call the
NG boson as {\it flaxion}~\cite{Ema:2016ops}.  (See
Refs.~\cite{Wilczek:1982rv,Geng:1988nc,Berezhiani:1989fp,Babu:1992cu,Albrecht:2010xh,Fong:2013sba,Ahn:2014gva,Ahn:2016hbn,Celis:2014iua,Celis:2014jua,Baryakhtar:2013wy}
for earlier related works, and also Refs.~\cite{Higaki:2016yeh,Suematsu:2017hki,Arias-Aragon:2017eww,Bjorkeroth:2017tsz} for more recent works.)  Moreover, it was shown~\cite{Ema:2016ops}
that the radial component of the U(1) breaking field can cause the
inflation, and the cosmological baryon asymmetry is naturally
explained by the decay of the right-handed neutrino via thermal
leptogenesis~\cite{Fukugita:1986hr}.

In this paper, we consider supersymmetric (SUSY) extensions of the flaxion scenario.
In the original (non-SUSY) flaxion scenario, the U(1) breaking field $\phi$ and the Standard Model Higgs field $H$ can have a direct coupling as
\begin{align}
{\cal L} &= -\lambda_{\phi H} |\phi|^2 |H|^2\,.
\label{eq:phi2H2}
\end{align}
After the U(1) symmetry breaking, the coupling induces a large correction to the Higgs mass, $\delta m_H^2 = \lambda_{\phi H} |\langle\phi\rangle |^2$. This must be cancelled by the bare Higgs mass term, requiring a severe fine-tuning between the model parameters. Note that, even if the quartic coupling in \eqref{eq:phi2H2} is set to vanish at the cut-off scale, it is induced via radiative corrections with quark/lepton loops.
Of course, even without a coupling like (\ref{eq:phi2H2}), the Higgs boson mass is quadratically divergent
due to gauge, Yukawa and Higgs self couplings and hence a severe fine-tuning is required in the non-SUSY model.
In SUSY models, the Higgs mass is protected against these large quadratic corrections, and such a fine-tuning can be avoided.\footnote{Although there is still a little hierarchy/fine-tuning originated from the difference between the electroweak scale and the (heavy) SUSY particle masses, we do not address this issue in this work.}

Compared to the non-SUSY case, there are several new ingredients in SUSY flaxion models. We discuss CP and flavor violations due to the SUSY particles, the effects of $R$-parity violation, and the cosmological problems of the  gravitino and the axino. Moreover, in the SUSY flaxion model, there is a light scalar field which corresponds to the flat direction of the superpotential, {\it sflaxion}. Its mass is expected to be of the order of the gravitino mass, and the cosmological evolution of such a light scalar field may spoil the standard thermal history.  We investigate the cosmological evolution of the sflaxion field, and show that there is a viable parameter space in which the leptogenesis can generate a sufficient amount of the baryon asymmetry.
We also propose an attractor-like inflationary model  in which the flaxion multiplet contains the inflaton field.

This paper is organized as follows. In Sec.~\ref{sec:SUSY_Flaxion}, we
present a SUSY flaxion model and investigate its implications for
phenomenology and cosmology. The model is introduced in
Sec.~\ref{sec:model}, which explains the mass hierarchies and mixings
of the quarks and leptons. In Sec.~\ref{sec:flaxion_pheno},
phenomenology of the light flaxion field is briefly summarized, which
is essentially the same as the non-SUSY case. Flavor and CP violations
due to the SUSY particles are discussed in Sec.~\ref{sec:SUSYcp}, and
the $R$-parity violations are examined in Sec.~\ref{sec:Rparity}. In
Sec.~\ref{sec:gravitino}, we investigate whether the leptogenesis
works in the present model, taking account of the cosmological
gravitino/axino problems. The cosmological evolution of the sflaxion
field is studied in Sec.~\ref{sec:oscillation}. In
Sec.~\ref{sec:inflation}, we propose an attractor-like inflation model
in the SUSY flaxion scenario, and show that a consistent and viable
cosmological history can be obtained. We summarize our findings in
Sec.~\ref{sec:summary}.

\section{Supersymmetric Flaxion} 
\label{sec:SUSY_Flaxion}

\subsection{Model}
\label{sec:model}

We introduce a global U(1) symmetry, U(1)$_{\text{F}}$, which is
broken by the vacuum expectation value (VEV) of the flaxion superfield
$\phi$ with a charge $q_\phi=-1$.  The U(1)$_{\text{F}}$-invariant
superpotential for the fields in the minimal SUSY standard model
(MSSM) and the right-handed neutrinos are given by
\begin{align}
W_{\text{MSSM}+N} = &\,
y_{ij}^u \lrfp{\phi}{M}{n_{ij}^u} Q_i \bar{U}_j H_u
+
y_{ij}^d \lrfp{\phi}{M}{n_{ij}^d} Q_i \bar{D}_j H_d
\nonumber\\
&
+
y_{ij}^\ell \lrfp{\phi}{M}{n_{ij}^\ell} L_i \bar{E}_j H_d
+
y_{i\alpha}^\nu \lrfp{\phi}{M}{n_{i\alpha}^\nu} L_i N_\alpha H_u
\nonumber\\
&
+
\frac{1}{2}
y_{\alpha\beta}^N \lrfp{\phi}{M}{n_{\alpha\beta}^N} M N_\alpha N_\beta
+
y^\mu \lrfp{\phi}{M}{n^\mu} M H_u H_d
\label{eq:W_MSSM}
\end{align}
with
\begin{align}
& 
n^u_{ij} = q_{Q_i} + q_{\bar{U}_j} + q_{H_u},\quad
n^d_{ij} = q_{Q_i} + q_{\bar{D}_j} + q_{H_d},
\nonumber \\
& 
n^\ell_{ij} = q_{L_i} + q_{\bar{E}_j} + q_{H_d},\quad
n^\nu_{i\alpha} = q_{L_i} + q_{N_\alpha} + q_{H_u},
\nonumber \\
& 
n^N_{\alpha\beta} = q_{N_\alpha} + q_{N_\beta},\qquad
n^\mu = q_{H_u} + q_{H_d}.
\label{eq:exponents_n}
\end{align}
Here, 
$Q_i, \bar{U}_i, \bar{D}_i, L_i, \bar{E}_i$ ($i=1$--$3$), $H_u$ and $H_d$ denote the superfields for the left-handed quarks, right-handed up- and down-type quarks, left-handed leptons, right-handed charged leptons, and up- and down-type Higgs fields, respectively. We also introduce two or three right-handed neutrinos $N_\alpha$ ($\alpha=1,2$ or $\alpha=1$--$3$). $M$ is a mass scale corresponding to the cut-off scale of this model, and
$q_X$ denotes the charge of the superfield $X$.\footnote{The sign convention is different from~\cite{Ema:2016ops}.}
Note that the exponents $n^{u,d,\ell,\nu,N,\mu}$ must be integers in order to allow the corresponding couplings, but $q_X$ are not necessarily integers.
For simplicity, we assume that charges are assigned such that all the entries of $n^{u,d,\ell,\nu,N,\mu}$ in \eqref{eq:exponents_n} are non-negative.\footnote{If some of them are negative, the corresponding coupling can be induced by using the field $\bar{\phi}$ with a charge $q_{\bar{\phi}}=+1$, which is introduced below.}

We assume that all the dimensionless couplings in the superpotential \eqref{eq:W_MSSM} are of order one, 
$y_{ij}^u, y_{ij}^d, y_{ij}^\ell, y_{i\alpha}^\nu, y_{\alpha\beta}^N, y^\mu \sim {\cal O}(1)$.
After the flaxion field develops a VEV $\vev{\phi}$, 
the hierarchies in the Yukawa couplings of the quarks and leptons are induced by the powers of the ratio
\begin{align}
\epsilon\equiv \frac{\vev{\phi}}{M} < 1,
\end{align}
as
\begin{align}
y^{u,\text{eff}}_{ij} = y^{u}_{ij} \epsilon^{n^u_{ij}},\;
y^{d,\text{eff}}_{ij} = y^{d}_{ij} \epsilon^{n^d_{ij}},\;
y^{\ell,\text{eff}}_{ij} = y^{\ell}_{ij} \epsilon^{n^\ell_{ij}},\;
y^{\nu,\text{eff}}_{i\alpha} = y^{\nu}_{i\alpha} \epsilon^{n^\nu_{i\alpha}}.
\end{align}
We further assume that the scales of the right-handed neutrino masses and the $\mu$-term are also induced by the powers of $\epsilon$;
\begin{align}
m_{N,\alpha\beta} = y^N_{\alpha\beta} \epsilon^{n^N_{\alpha\beta}} M,
\quad
\mu = y^\mu \epsilon^{n^\mu} M.
\end{align}
For instance, the masses of the quarks and leptons as well as their mixings are obtained by taking $\epsilon \simeq 0.23$ and the following values of $n^{u,d,\ell}_{ij}$ (cf.~\cite{Ema:2016ops})
\begin{align}
	&n_{ij}^u=\begin{pmatrix}
		8 & 4 & 3 \\
		7 & 3 & 2 \\
		5 & 1 & 0 
	\end{pmatrix},\;
	n_{ij}^d=\begin{pmatrix}
		7-\Delta_d & 6-\Delta_d & 6-\Delta_d \\
		6-\Delta_d & 5-\Delta_d & 5-\Delta_d \\
		4-\Delta_d & 3-\Delta_d & 3-\Delta_d 
	\end{pmatrix},\;
	n_{ij}^\ell=\begin{pmatrix}
		9-\Delta_d & 6-\Delta_d & 4-\Delta_d \\
		8-\Delta_d & 5-\Delta_d & 3-\Delta_d \\
		8-\Delta_d & 5-\Delta_d & 3-\Delta_d 
	\end{pmatrix},
	\label{eq:charge_matrix}
\end{align}
where $\Delta_d$ is an integer which depends on $\tan\beta =
\vev{H_u}/\vev{H_d}$ as $1/\tan\beta \sim \epsilon^{\Delta_d}$.  The
above $n^{u,d,\ell}_{ij}$ can be obtained by the charge assignments
shown in Table.~\ref{tab:charges}. Here, we take $q_{H_u}=0$ by using
the U(1)$_Y$ rotation. The parameters $\ell, b, h$ as well as
$q_{N_\alpha}$ are not determined at this stage.  In the following, we
take $\epsilon=0.23$, $\Delta_d=1$ ($\tan\beta\simeq 4$), and the
matrices in \eqref{eq:charge_matrix} as representative values,
although the main conclusions do not depend on the details of the
charges assignments.
\begin{table}[t]
\begin{center}
\begin{tabular}{|r|ccccc|cc|c|c|}
\hline
$q_X$ & $Q_i$ & $\bar{U}_i$ & $\bar{D}_i$ & $L_i$ & $\bar{E}_i$ & $H_u$ & $H_d$ & $N_\alpha$ & $\phi$
\\ \hline
$i=1$ & $b+3$ & $-b+5$ & $-b-h+4-\Delta_d$ & $\ell+1$ & $-\ell-h+8-\Delta_d$ & $0$ & $h$ & $q_{N_\alpha}$ & $-1$
\\
$2$ & $b+2$ & $-b+1$ & $-b-h+3-\Delta_d$ & $\ell$    & $-\ell-h+5-\Delta_d$ &&&&
\\
$3$ & $b$    & $-b$    & $-b-h+3-\Delta_d$ & $\ell$    & $-\ell-h+3-\Delta_d$ &&&&
\\ \hline
\end{tabular}
\end{center}
\caption{An example of charge assignments.  Here, 
  $b$, $\ell$, $h$, $\Delta_d$, and $q_{N_\alpha}$ are parameters (see text).}
\label{tab:charges}
\end{table}

The scale of the $\mu$-term is controlled by the parameter $h$, \textit{i.e.},
the U(1)$_{\text{F}}$ charge of the $H_d$ field. For instance,
$\mu\simeq {\cal O}(1\text{--}10^3)~\text{TeV}$ is obtained by taking $n^\mu = h \simeq 11\hf\hf16$ for $M\simeq 10^{13}~\text{GeV}$ and $\epsilon = 0.23$.

In the neutrino sector, after integrating out the heavy right-handed neutrinos, the neutrino masses are induced by the seesaw mechanism~\cite{Yanagida:1979as,GellMann:1980vs,Minkowski:1977sc}. The neutrino mass matrix at low energy is given by
\begin{align}
m_{\nu,ij} = \kappa^{\nu}_{ij} \epsilon^{(q_{L_i}+q_{L_j}+2q_{H_u})} \frac{\vev{H_u}^2}{M},
\end{align}
where the ${\cal O}(1)$ coefficients $\kappa^\nu_{ij}$ are given by $\kappa^\nu = -y^\nu (y^N)^{-1} (y^\nu)^T$.
Note that $m_{\nu,ij}$ is independent of the charges of the right-handed neutrinos $q_{N_\alpha}$. 
In particular, the largest neutrino mass is given by
\begin{align}
    m_{\nu,3}\sim \epsilon^{2\ell}\frac{\vev{H_u}^2}{M}.\label{eq:charge_nu_mass}
\end{align}
For instance, the desired value $m_{\nu,3}\simeq 0.05~\text{eV}$ is obtained by $2\ell\simeq 2, 3$ for $M\simeq 10^{13}~\text{GeV}$.

Finally, as the U(1)$_{\text{F}}$ breaking sector, we assume the
following superpotential and K\"ahler potential;
\begin{align}
W_{\text{flaxion}} &=\lambda S(\phi \bar{\phi}-f^2)\,,
\label{eq:Wflaxion}
\\
K_{\text{flaxion}} &=|\phi|^2 + |\bar\phi|^2 + |S|^2\,
\label{eq:Kflaxion}
\end{align}
where $\bar{\phi}$ is a superfield with charge $q_{\bar{\phi}}=+1$.
After the global U(1)$_{\text{F}}$ symmetry is spontaneously broken,
the scalar components of the superfields $\phi$, $\bar\phi$ can be
expanded as
\begin{align}
\phi &= \vev{\phi} + \frac{1}{\sqrt{2}}(s+i\theta) 
\simeq \vev{\phi} + \frac{\sin\alpha}{\sqrt{2}}(\chi + i\eta) + \frac{\cos\alpha}{\sqrt{2}}(\sigma - ia),
\nonumber \\
\bar\phi &= \vev{\bar\phi} + \frac{1}{\sqrt{2}}(\bar{s}+i\bar\theta) 
\simeq \vev{\bar\phi} +\frac{\cos\alpha}{\sqrt{2}}(\chi + i\eta) - \frac{\sin\alpha}{\sqrt{2}}(\sigma - ia),
\label{eq:phi_phibar_a_sigma}
\end{align}
where $\vev{\phi}\vev{\bar\phi}=f^2$, $\tan\alpha=\vev{\bar{\phi}}/\vev{\phi}$, 
and we have neglected the corrections from the soft SUSY breaking terms.
In the following discussion, we assume $\vev{\phi}=\vev{\bar\phi}=f=$ real, for simplicity.
The field $a$ is the NG boson associated with the broken U(1)$_{\text{F}}$, which is identified with the QCD axion that solves the strong CP problem~\cite{Ema:2016ops,Calibbi:2016hwq} (see Sec.~\ref{sec:flaxion_pheno}). We call $a$ as {\it flaxion}~\cite{Ema:2016ops}. The field $\sigma$ corresponds to the flat direction of the superpotential \eqref{eq:Wflaxion}, which can be called as {\it sflaxion}. It receives a soft mass of the order of the gravitino mass, $m_\sigma \simeq {\cal O}(m_{3/2})$, after SUSY breaking. Its cosmological implication will be discussed in Secs.~\ref{sec:oscillation} and \ref{sec:alpha_inf_sflaxion}.
The orthogonal fields, $\chi$ and $\eta$, receive large masses of ${\cal O}(\lambda f)$.
In Sec.~\ref{sec:inflation}, we will discuss an inflation model where the inflaton corresponds to the $\chi$ direction.

\subsection{Solution to the strong-CP problem, flaxion dark matter, and flavor-changing signatures}
\label{sec:flaxion_pheno}
In this subsection, we briefly review some of the phenomenology of the flaxion $a$ that are essentially the same as the non-SUSY case~\cite{Ema:2016ops,Calibbi:2016hwq}. 

\begin{itemize}

\item {\bf Flaxion as QCD axion:}
Since the U(1)$_{\text{F}}$ is anomalous under the QCD, the interactions between the flaxion $a$ and the quarks induce an effective flaxion-gluon-gluon coupling as
\begin{align}
{\cal L}&=\frac{g_s^2}{32\pi^2}\frac{a}{f_a}G^a_{\mu\nu} \widetilde{G}^{a\mu\nu},
\label{eq:aGG}
\end{align}
where
\begin{align}
f_a = \frac{2\vev{\phi}}{N_{\text{DW}}} = \frac{2\epsilon M}{N_{\text{DW}}}\,,
\label{eq:fa_phi_M_1}
\end{align}
with the domain-wall (DW) number $N_{\text{DW}}=|\text{Tr}(n^u+n^d)-3h|$.\footnote{The coefficient in Eq.~\eqref{eq:fa_phi_M_1} is different from that in Ref.~\cite{Ema:2016ops} by a factor of $\sqrt{2}$, because of the presence of the $\bar{\phi}$ field. See Eq.~\eqref{eq:phi_phibar_a_sigma}.}
For the charge assignments in Table~\ref{tab:charges} with $\Delta_d=1$, we have $N_{\text{DW}}=|23-3h|$, which takes a wide range of values depending on $h$ (or $n^\mu$). For example, $M\simeq 10^{13}~\text{GeV}$ and $\mu\simeq 10^3~\text{TeV}$ lead to $n^\mu=11$ and $N_{\text{DW}}= 10$, resulting in
\begin{align}
f_a : \vev{\phi} : M \simeq 1 : 5 : 22.
\end{align}
As in the ordinary QCD axion, below the QCD confinement scale, the interaction \eqref{eq:aGG} leads to a potential of the flaxion $a$ that solves the strong CP problem via the PQ mechanism.
The flaxion mass is given by~\cite{Kim:1986ax}
\begin{align}
m_a\simeq 6\times 10^{-6}\text{eV}\left(\frac{10^{12}\,\text{GeV}}{f_a}\right).
\end{align}

\item {\bf Flaxion dark matter:}
Most of the low-energy phenomenology and cosmology of the flaxion $a$ are similar to the ordinary QCD axion.  In particular, its coherent oscillation in the present universe can play the role of the cold dark matter. Its present energy density is given by~\cite{Turner:1985si}
\begin{align}
\Omega_a h^2 \simeq 0.18\; \Theta_i^2 \left(\frac{f_a}{10^{12}\,\text{GeV}}\right)^{1.19},\label{flaxion_DM}
\end{align}
where $\Theta_i$ is the initial misalignment angle ($0\le \Theta_i < 2\pi$). The flaxion can be the dominant component of the dark matter for $f_a\gtrsim {\cal O}(10^{11})~\text{GeV}$. For $N_{\text{DW}}= 10$ and $\epsilon \simeq 0.23$, this corresponds to $\vev{\phi}, M \gtrsim  {\cal O}(10^{12})~\text{GeV}$.

\item {\bf Flavor changing signature of the flaxion:} A unique
  signature of the flaxion scenario is its flavor changing
  signature~\cite{Ema:2016ops,Calibbi:2016hwq}. Currently the most
  stringent constraint on the PQ scale $f_a$ comes from the process
  $K^+ \to \pi^+ a$, which is induced by the following interaction
  \begin{align}
    -\mathcal L &\supset i 
    \frac{(\kappa^d_{\text{AH}})_{21}}{2\langle\phi\rangle}
    a \overline {s} d ,
  \end{align}
  where the quark fields are the mass eigenstates, and $(\kappa^d_{\text{AH}})_{21}$ is determined by the
  U(1)$_{\text{F}}$ charges of quarks and the ${\cal O}(1)$ couplings
  $y^d_{ij}$. (For the definition of $(\kappa^d_{\text{AH}})_{21}$, see Ref.~\cite{Ema:2016ops}.)
  For the charge assignment in Table~\ref{tab:charges},
  $|(\kappa^d_{\text{AH}})_{21}|/m_s\sim {\cal O}(\epsilon)$ (or larger), with
  $m_s$ being the strange quark mass. The decay rate is given by
  \begin{align}
    \text{Br}(K^+\to \pi^+ a)\simeq 
    4\times 10^{-9}
    \left(\frac{10^{10}\,\text{GeV}}{f_a}\right)^2
    \left(\frac{10}{N_{\text{DW}}}\right)^2
    \left| \frac{ (\kappa^d_{\text{AH}})_{21} }{m_s} \right|^2.
  \end{align}
  The current experimental bound $\text{Br}(K^+\to \pi^+ a)\lesssim
  7.3\times 10^{-11}$~\cite{Adler:2008zza} leads to a bound
  \begin{align}
    f_a\gtrsim 7\times 10^{10}~\text{GeV}\;
    \left(\frac{10}{N_{\text{DW}}}\right)
    \left(\frac{|(\kappa^d_{\text{AH}})_{21}|}{m_s}\right).
  \end{align}
  The constraint may be improved by the NA62 experiment
  \cite{Moulson:2016dul} in near future.  
  \end{itemize}

\subsection{Flavor and CP violations due to superparticles}
\label{sec:SUSYcp}

In the present model, the soft SUSY breaking mass terms of the
sfermions are expected to be controlled by the U(1)$_{\rm F}$ symmetry,
assuming that there exist all interactions allowed by the U(1)$_\mathrm{F}$
symmetry between observable and
SUSY breaking sector fields.
Such a situation is potentially problematic because the
soft SUSY breaking parameters may induce new sources of CP and flavor
violating processes.  One solution to such a problem is to increase the
masses of the superparticles.  Here, we discuss how much the
masses of superparticles should be, adopting our canonical choice of
the U(1)$_{\rm F}$ charge assignments given in
Table~\ref{tab:charges}.

Let us parametrize the soft SUSY breaking mass terms of sfermions as
\begin{align}
  {\cal L}_{\rm mass} = m_{\tilde{f}}^2
  \sum_{F=Q,\bar{U},\bar{D},L,\bar{E}} 
  \Delta_{\tilde{F}_{ij}}
  \tilde{F}_i^\dagger \tilde{F}_j,
\end{align}
where $\Delta_{\tilde{F}_{ij}}$'s are dimensionless parameters.  In
addition, $m_{\tilde{f}}$ is a parameter with the mass dimension of
$1$.  Hereafter, we assume that the soft SUSY breaking sfermion masses
are generated by the Planck-suppressed K\"ahler interactions between
sfermions and SUSY breaking fields (like in the case of gravity
mediated SUSY breaking scenario), and $m_{\tilde{f}}$ is of the order
of the gravitino mass.  Then, because the sfermions have the same
U(1)$_{\rm F}$ charges as their superpartners, we expect
\begin{align}
  \Delta_{\tilde{F}_{ij}} \sim \mathcal{O}(\epsilon^{|q_{F_i}-q_{F_j}|}).
\end{align}
The off-diagonal elements of $\Delta_{\tilde{F}_{ij}}$ are not only
non-vanishing in the present model but also complex in general, which
will become important in the following discussion.

With the charge assignment given in Table~\ref{tab:charges}, the most
significant constraint on the mass scale of superparticles is obtained
from the $K^0$-$\bar{K}^0$ mixing parameter (so-called $\epsilon_K$).
Assuming that the gluino mass is $\sim m_{\tilde{f}}$, the SUSY
contribution to the $\epsilon_K$ parameter is estimated as
\cite{Moroi:2013sfa}\footnote
{The following argument is qualitatively unchanged even if the gluino 
mass is order-of-magnitude smaller than $m_{\tilde{f}}$, as suggested
by the anomaly-mediation scenario.}
\begin{align}
  \epsilon_K^{\rm (SUSY)} \sim 
  2 \times 10^{-3} \times
  \left( \frac{m_{\tilde{f}}}{1000\,{\rm TeV}} \right)^{-2}
  \mbox{Im}
  \left[
    \left( \frac{\Delta_{\tilde{Q}_{12}}}{0.23} \right)
    \left( \frac{\Delta_{\tilde{\bar{D}}_{12}}^*}{0.23} \right)
  \right].
\end{align}
Comparing with the experimental value of $\epsilon_K$ (\textit{i.e.},
$\epsilon_K^{\rm (exp)}\simeq 1.596\times 10^{-3}$
\cite{Patrignani:2016xqp}), the sfermion mass scale is required to be
$\sim 10^3\ {\rm TeV}$ or larger if the CP is maximally violated.

Other check points are the lepton-flavor violating processes.  With
currently available experimental data, $\mu\rightarrow e\gamma$ gives
the most stringent constraint on the mass scale of superparticles.
Assuming that the masses of all the superparticles are of the same
order,\footnote
{If the gaugino masses (which we denote $M_{\rm gaugino}$ in this
  footnote) are significantly lighter than the sfermion and Higgsino
  masses, the branching ratio is suppressed by the factor of $\sim
  O(M_{\rm gaugino}^2/m_{\tilde{f}}^2)$.  This is the case, for
  example, in the anomaly-mediation scenario.}
the branching ratio for the process $\mu\rightarrow e\gamma$ is
estimated as
\begin{align}
  {\rm Br} (\mu\rightarrow e\gamma)
  \sim 3 \times 10^{-12}
  \left( \frac{\tan\beta}{10} \right)^{2}
  \left( \frac{m_{\tilde{f}}}{10\, {\rm TeV}} \right)^{-4}
  \left| \frac{\Delta_{\tilde{L}_{12}}}{0.23} \right|^2.
\end{align}
Here, we have used the fact that, with the charge assignment given in
Table~\ref{tab:charges}, $|\Delta_{\tilde{L}_{12}}|$ is expected to be
larger than $|\Delta_{\tilde{E}_{12}}|$.  Experimentally, ${\rm Br}
(\mu\rightarrow e\gamma)<4.2\times 10^{-13}$ \cite{TheMEG:2016wtm},
which requires $m_{\tilde{f}}$ be of $\mathcal{O}(10)\ {\rm TeV}$ or larger for
$\tan\beta\sim \mathcal{O}(10)$.

\subsection{$R$-parity violating terms}
\label{sec:Rparity}

So far we have discussed only the $R$-parity conserving interactions in the superpotential \eqref{eq:W_MSSM}. However, depending on the U(1)$_\mathrm{F}$
charges, the $R$-parity violating terms, 
$\bar{U}_i \bar{D}_j \bar{D}_k$, $L_i L_j \bar{E}_k$, $L_i Q_j \bar{D}_k$ and $L_i H_u$
can be induced after the flaxion gets the VEV.
In the present work, instead of imposing the $R$-parity as an additional symmetry, we can assume that those interactions are also controlled by the U(1)$_{\text{F}}$ symmetry.
There are severe constraints on those couplings, whereas the $R$-parity violation can also circumvent the gravitino/axino constraints as we discuss in the next subsection. 

Let us start from the lepton-number violating operators, $L_i L_j \bar{E}_k$, $L_i Q_j \bar{D}_k$ and $L_i H_u$. Their charges are all given by $q = \ell + n$ with $n \in {\bf Z}$. 
In order to generate the neutrino masses, the lepton charge $\ell$ (and $q_{N_\alpha}$) must be integer or half-integer, \textit{i.e.}, $\ell=n$ or $\ell=n+1/2$ ($n \in {\bf Z}$). 
If $\ell=n+1/2$, the couplings $L_i L_j \bar{E}_k$, $L_i Q_j \bar{D}_k$ and $L_i H_u$ are forbidden, whereas, if $\ell=n$, those couplings can be induced by the VEV of $\phi$ or $\bar{\phi}$.

In the case of $\ell\in {\bf Z}$, the lepton-number violating operators are induced.
In particular, the bilinear terms would be induced as
\begin{align}
W_{L H_u} = 
y^{\mu'}_3 \epsilon^{|\ell |} M L_3 H_u 
+ y^{\mu'}_2 \epsilon^{|\ell |} M L_2 H_u
+ y^{\mu'}_1 \epsilon^{|\ell +1|} M L_1 H_u,
\end{align}
where $y^{\mu'}_i$ are coefficients of order unity, and
$\epsilon^{|\ell |}, \epsilon^{|\ell +1|}$ are induced by the VEV of
$\phi$ or $\bar{\phi}$. The strongest bound on these couplings comes
from the cosmology.  Assuming that the baryon asymmetry is generated
before the electroweak phase transition, the lepton number violating
processes induced by the $R$-parity breaking couplings (together with
the sphaleron process~\cite{Kuzmin:1985mm}) would wash out the
existing baryon
asymmetry~\cite{Campbell:1990fa,Fischler:1990gn,Dreiner:1992vm,Endo:2009cv}. The
constraint reads $y^{\mu'}_{2,3} \epsilon^{|\ell |} M/\mu \lesssim
{\cal O}(10^{-6}) (m_{\tilde{f}}/{\rm TeV})^{1/2}$~\cite{Endo:2009cv},
corresponding to $|\ell |\gtrsim 18\hf\hf 25$ for $M\simeq 10^{13}~\text{GeV}$ and $\mu\sim
m_{\tilde{f}}\sim (1\hf\hf10^3)~\text{TeV}$.  Such a large $|\ell |$ is clearly
inconsistent with the requirement from the neutrino mass in
Eq.~\eqref{eq:charge_nu_mass}.  Thus, we have to take $\ell=n+1/2$ with $n\in {\bf Z}$ or impose lepton parity as an additional symmetry.\footnote{The
  bilinear $R$-parity violation itself can generate the neutrino
  masses~\cite{Hall:1983id}, but the leptogenesis does not work in that case.}

Next, we consider the baryon-number violating operators $\bar{U}_i \bar{D}_j \bar{D}_k$. The charges of those operators are $q=-3b+n$ ($n\in {\bf Z}$). Thus, those couplings are forbidden for $3b \notin {\bf Z}$. For $3b\in {\bf Z}$, the interactions $\bar{U}_i \bar{D}_j \bar{D}_k$ are induced,
\begin{align}
W_{\bar{U}\bar{D}\bar{D}} = \frac{1}{2}\lambda''_{ijk} \bar{U}_i \bar{D}_j \bar{D}_k
=
\frac{1}{2} y''_{ijk} \epsilon^{|q_{\bar{U}_i}+q_{\bar{D}_j}+q_{\bar{D}_k}|} \bar{U}_i \bar{D}_j \bar{D}_k
\end{align}
with the largest coupling given by (cf.~Table.\ref{tab:charges})
\begin{align}
\lambda''_{\text{max}} \sim \epsilon^{n''_{\text{min}}},
\qquad
n''_{\text{min}} = \min_{n=6,7,8,11,12} |-3b-2h+n-2\Delta_d|.
\end{align}
Again, the severest constraint comes from the baryon washout, which is
given by $\lambda''_{\text{max}} \lesssim {\cal
  O}(10^{-6})(m_{\tilde{f}}/{\rm TeV})^{1/2}$~\cite{Endo:2009cv},
corresponding to $n''_{\text{min}} \gtrsim 7\hf\hf 9$ for
$m_{\tilde{f}}\sim (1\hf\hf10^3)~\text{TeV}$. Implications of the $R$-parity
conservation/violation are discussed with the gravitino/axino problem
in the next subsection.

Before closing this subsection, let us briefly mention the proton
decay. In supersymmetric models, dimension-5 operators in the
superpotential, such as $W\sim QQQL$ and
$\bar{U}\bar{U}\bar{D}\bar{E}$, may cause a rapid proton decay, 
and therefore the coefficients of those operators are
severely constrained. In the present SUSY flaxion scenario, those
operators have U(1)$_\mathrm{F}$ charges of $q=3b+\ell+n$ ($n\in {\bf Z}$), and are
forbidden for $3b+\ell \notin {\bf Z}$. Otherwise, they can be prohibited by lepton parity.

\subsection{Leptogenesis and gravitino/axino problem}  
\label{sec:gravitino}

In the present universe, non-zero baryon asymmetry is observed. If the reheating temperature $T_R$ is high enough, thermal leptogenesis~\cite{Fukugita:1986hr} can be one of the promising baryogenesis scenarios. Noting that the present model favors strong washout regime, the baryon-to-entropy ratio obtained through thermal leptogenesis for $m_{N_1} \ll m_{N_{2(3)}}$ is calculated as~\cite{Ema:2016ops}
\begin{align}
	\frac{n_B}{s} \lesssim 6\times 10^{-11}~ \gamma \left(\frac{m_{N_1}}{10^{11}\, \text{GeV}}\right),
	\label{baryon_asymmetry_susy}
\end{align}
where the right-hand side has been evaluated for the maximal CP asymmetry.
Here, we have taken account of the enhancement of the baryon asymmetry by a factor of $\sim 1.4$ as compared to non-SUSY cases \cite{Davidson:2008bu}. Furthermore, we have introduced the dilution factor $\gamma$ ($\le 1$), which represents the effect of dilution due to late-time entropy production and its decent definition will be given in the next subsection. From the expression, if $\gamma\simeq 1$, the observed baryon asymmetry $n_B/s\simeq 9\times 10^{-11}$ can be obtained for $m_{N_1}\sim \mathcal{O}(10^{11})\, \text{GeV}$. In the next subsection, we will investigate the influence of $\gamma$ on thermal leptogenesis.

On the other hand, in SUSY, high $T_R$ necessary for thermal leptogenesis is not favored from the perspective of the gravitino production~\cite{Khlopov:1984pf,Ellis:1984eq}.
For high $T_R$, gravitinos produced by thermal scatterings can overclose the universe if it is stable~\cite{Moroi:1993mb}. 
If the gravitino is unstable and has a lifetime longer than ${\cal O}(1)$~sec, it may spoil the success of the big-bang nucleosynthesis (BBN)~\cite{Kawasaki:2017bqm}. 
Thus let us assume that the gravitino is unstable and decays before the BBN, which is the case for $m_{3/2}\gtrsim \mathcal{O}(10)$ TeV. Even in this case, the lightest SUSY particle (LSP) produced from the gravitino decay may overclose the universe for high $T_R$. Fortunately, in the present model, we can find a way out of this dilemma for thermal leptogenesis in SUSY: the $R$-parity violation, leading to the LSP decay. As discussed in Sec.~\ref{sec:Rparity}, we can realize a situation, in terms of the charge assignment ($\ell-1/2,~3b \in {\bf Z}$ or $\ell,~3b \in {\bf Z}$ with lepton parity), where the $\bar{U}_i\bar{D}_j\bar{D}_k$ terms can exist with the proton decay forbidden. 

Let us investigate whether any acceptable charge assignments actually exist or not.
First, the gravitino mass must be larger than $\mathcal{O}(10)$ TeV so that the gravitino can decay well before the BBN. Furthermore, the LSPs must also decay before the BBN. Otherwise, the decay products of the LSPs spoil the successful BBN. 
This requires the LSP lifetime to be shorter than $\mathcal{O}(0.1)$ sec~\cite{Kawasaki:2017bqm}.\footnote{This upper bound on the lifetime is the strongest one. It can be somewhat relaxed depending on the abundance and the decay mode of the LSP.}
Let us first consider the case that the LSP is the Bino. The decay rate of the Bino is given by~\cite{Baltz:1997gd}
\begin{align}
	\Gamma_{\tilde{B}^0} = \frac{g'^2}{256\pi^3}\frac{m_{\tilde{B}^0}^5}{m_{\tilde{q}}^4}\sum_{i,j,k\,(j<k)}|\lambda''_{ijk}|^2 \simeq \frac{g'^2}{256\pi^3}\frac{m_{\tilde{B}^0}^5}{m_{\tilde{q}}^4}|\lambda''_\text{max}|^2,
\end{align}
where $g'$, $m_{\tilde{q}}$, and $m_{\tilde{B}^0}$ denote the U(1)$_Y$ gauge coupling constant, the squark mass, and the Bino mass, respectively. Thus, the coupling constant must satisfy
\begin{align}
	|\lambda''_\text{max}| \gtrsim 6\times 10^{-9}\left(\frac{m_{\tilde{q}}}{10\, \text{TeV}}
\right)^2 \left(\frac{m_{\tilde{B}^0}}{1\, \rm{TeV}}\right)^{-5/2}.
\end{align}
Together with the upper bound from the baryon washout mentioned in the previous subsection, $\lambda''_\text{max}$ must lie in some limited range. For $m_{\tilde{B}^0}\simeq \mathcal{O}(0.1)m_{\tilde{q}}$ and $m_{\tilde{q}}\simeq m_{3/2}\gtrsim \mathcal{O}(10)\, \text{TeV}$, we can easily find possible charge assignments to realize the allowed values of $\lambda''_\text{max}$. 

Next, we consider the case that the Wino is the LSP and the squarks
and sleptons are much heavier. Without singlet fields in the SUSY
breaking sector, the dominant contribution to the gaugino masses can
be the anomaly mediated effect~\cite{Randall:1998uk,Giudice:1998xp},
and the gauginos can become significantly lighter than the gravitino. For
example, such a situation occurs in the framework of the heavy
sfermion scenarios~\cite{Hall:2011jd,Ibe:2011aa,Ibe:2012hu,Bhattacherjee:2012ed,Dudas:2012wi,Arvanitaki:2012ps,Hall:2012zp,ArkaniHamed:2012gw}, where the SM Higgs boson mass can be
explained and the most probable LSP candidate is the neutral
Wino. In that case, the decay rate of the Wino LSP scales as
$\Gamma_{\tilde{W}^0}\propto |\lambda_{323}''|^2 m_{\tilde{W}^0}\times
(m_t/m_{\tilde{q}})^2 (m_{\tilde{W}^0}/m_{\tilde{q}})^4(
(A_t+\mu\cot\beta)/m_{\tilde{q}})^2$, where $m_{\tilde W_0}$, $m_t$ and $A_t$ denote the Wino mass, top quark mass, and the
trilinear coupling constant among Higgs and stops, respectively. It is doubly
suppressed by the Wino mass and by the left-right mixing. We have
found that there is no allowed range of $\lambda''$ in this case.

The discussion so far is also true of the axino (or we may call it {\it flaxino}), the fermionic superpartner of the flaxion, which can cause problems similar to those of the gravitino. Although the axino mass is model-dependent, it can be comparable to the gravitino mass in general.\footnote{For our choice of the superpotential \eqref{eq:Wflaxion}, the axino mass is given by $\lambda \langle S\rangle$, which can be $\sim m_{3/2}$ after including the SUSY breaking effect.}
For simplicity here we assume $m_{\tilde{a}}\sim m_{3/2}$. The abundance of the thermally produced axino can be much larger than that of the gravitino~\cite{Chun:2011zd,Bae:2011jb,Bae:2011iw}. 
The $\mu$-term in \eqref{eq:W_MSSM} leads to the axino decay into a Higgsino and a Higgs boson, if kinematically allowed, which can be fast enough to make the axino decay well before the BBN and also before the axino domination of the energy density of the universe.\footnote{If the decay modes including the Higgsino are forbidden, the axino can dominate the energy density of the universe before it decays, which leads to entropy production and dilution.}
Since the produced LSP decays through the $R$-parity violating operators as mentioned above,
the U(1)$_{\rm F}$ charge assignments to circumvent the gravitino problem can make the axino cosmologically harmless as well.

\subsection{Cosmology of sflaxion}
\label{sec:oscillation}

One of the new important issues in the SUSY flaxion model is the existence of the sflaxion, 
which corresponds to the flat direction of the superpotential (cf. Sec.~\ref{sec:model}).
Its mass is expected to be of the order of the
gravitino mass, which is much smaller than the breaking scale of the
flavor symmetry $M$. 
After the reheating of the universe, coherent oscillation of the sflaxion can be induced through thermal effects, 
and its late-time decay can lead to large entropy production, diluting all kinds of abundances in the universe at that time, such as the baryon asymmetry, the gravitino relic density, and so on. 
If baryogenesis is to be realized by thermal leptogenesis, the dilution can be problematic.\footnote{Affleck-Dine mechanism~\cite{Affleck:1984fy,Dine:1995kz} may produce enough amount of baryon asymmetry even in the presence of sizable dilution~\cite{Kawasaki:2008jc}, although it is non-trivial whether this mechanism successfully works or not in the present model because of the U(1)$_{\rm F}$ symmetry.}
The sflaxion evolution and the amount of dilution are highly model-dependent and here we estimate them with a simplified setup. We reconsider this subject with an explicit model later in Sec.~\ref{sec:alpha_inf_sflaxion}.

During the radiation-dominated era after the reheating, the sflaxion
receives thermal effects from various fields in the thermal
bath that couple to $\phi$ 
(or $\bar{\phi}$)~\cite{Kawasaki:2010gv,Kawasaki:2011ym,Moroi:2012vu}.  
Here we focus on the one-loop effects, assuming that higher order ones are subdominant. The fields involved in the one-loop contributions are the ones with the bilinear terms in the superpotential \eqref{eq:W_MSSM}, \textit{i.e.}, the right-handed (s)neutrinos and the Higgs fields. The $\phi$-dependent effective
mass of such a field is expressed as
\begin{align}
	M_\text{eff}(\phi) \equiv yM\left( \frac{\phi}{M}\right)^{n},
\end{align}
where $n$ and $y$ are the corresponding exponent and the
$\mathcal{O}(1)$ coupling, respectively.  Then the contribution to the sflaxion potential 
at the one-loop level is given
by~\cite{Dolan:1973qd}
\begin{align}
	V_T(M_\text{eff}(\phi),T) \equiv \frac{T^4}{\pi^2}\int_0^\infty dz~z^2\ln\tanh\left( \frac{1}{2}\sqrt{z^2+\frac{M_\text{eff}^2(\phi)}{T^2}}\right)
	\underbrace{\simeq}_{T\gg M_\text{eff}} \frac{1}{8}y^2M^2T^2\left( \frac{\phi}{M}\right)^{2n},
	\label{V_T}
\end{align}
where we have added the contributions of the bosons and the fermions.
In the last expression, we have kept only a $\phi$-dependent leading
term for $T\gg M_\text{eff}$.  The behavior of this potential depends
on $n$; as $n$ increases, the potential becomes steeper with respect
to $\phi$ and is effective only at lower temperature.
Such thermal contributions may displace the sflaxion from the VEV at the present universe
and induce a coherent sflaxion oscillation, 
resulting in the dilution at its late-time decay.
In particular, if any non-zero $n^N_{\alpha\beta}$ exist, the contributions from the right-handed (s)neutrinos
are important in the present case. Otherwise, the contributions from the Higgs fields are important.\footnote{
	 The coherent sflaxion oscillation is less likely to be induced for
	 larger $n$ as we see below
	 (see Eq.~\eqref{no-oscillation}).
}

In order to quantify the amount of the dilution, 
we define the dilution factor $\gamma$ as
\begin{align}
	\gamma \equiv \frac{s_\text{before}}{s_\text{after}} \simeq 
	\min \left[ \frac{3}{4}T_\sigma\left(\frac{\rho_{\sigma,i}}{s_i}\right)^{-1}, 1\right],
	\label{dilution_factor}
\end{align}
where $s_\text{before (after)}$, $\rho_\sigma$, and $T_\sigma$ denote the entropy density just before (after) the sflaxion decay, the energy density of the sflaxion, and the sflaxion decay temperature, respectively. (Here and hereafter, quantities with a subscript $i$ are evaluated at the beginning of the sflaxion oscillation.) If the sflaxion decays dominantly into the Higgs bosons,\footnote{In the calculation, we include the four degrees of freedom in two Higgs doublets. For the decay into Higgsinos, the decay rate can be comparable. The decay mode into flaxions can be suppressed as long as the soft masses of $\phi$ and $\overline\phi$ are not so different~\cite{Chun:1995hc,Kawasaki:2007mk}.} the decay rate is given by
\begin{align}
	\Gamma_\sigma \simeq \frac{(n^\mu)^2}{8\pi}\frac{m_\sigma^3}{f^2}\left(\frac{\mu}{m_\sigma}\right)^4
	\simeq 7.5\times 10^{-7}\, \text{GeV} \left(\frac{n^\mu}{10}\right)^{2} \left(\frac{10^{13}\, \text{GeV}}{M}\right)^{2} \left(\frac{10^{3}\, \text{TeV}}{m_\sigma}\right) \left(\frac{\mu}{10^{3}\, \text{TeV}}\right)^{4}.
\end{align}
The decay temperature is then given by
\begin{align}
	T_\sigma &\simeq \left(\frac{90}{\pi^2g_*}\right)^{1/4}\sqrt{M_P\Gamma_\sigma}
	\simeq 6\times 10^5\, \text{GeV} \left(\frac{n^\mu}{10}\right) \left(\frac{10^{13}\, \text{GeV}}{M}\right) \left(\frac{10^{3}\, \text{TeV}}{m_\sigma}\right)^{1/2} \left(\frac{\mu}{10^{3}\, \text{TeV}}\right)^{2},
	\label{decay_temperature}
\end{align}
where $M_{\text{P}}\simeq 2.4\times 10^{18}~\text{GeV}$ is the reduced Planck scale and we have taken $g_*=228.75$.

\begin{figure}[t]
	\begin{minipage}{0.5\hsize}
		\begin{center}
		\includegraphics[width=8cm]{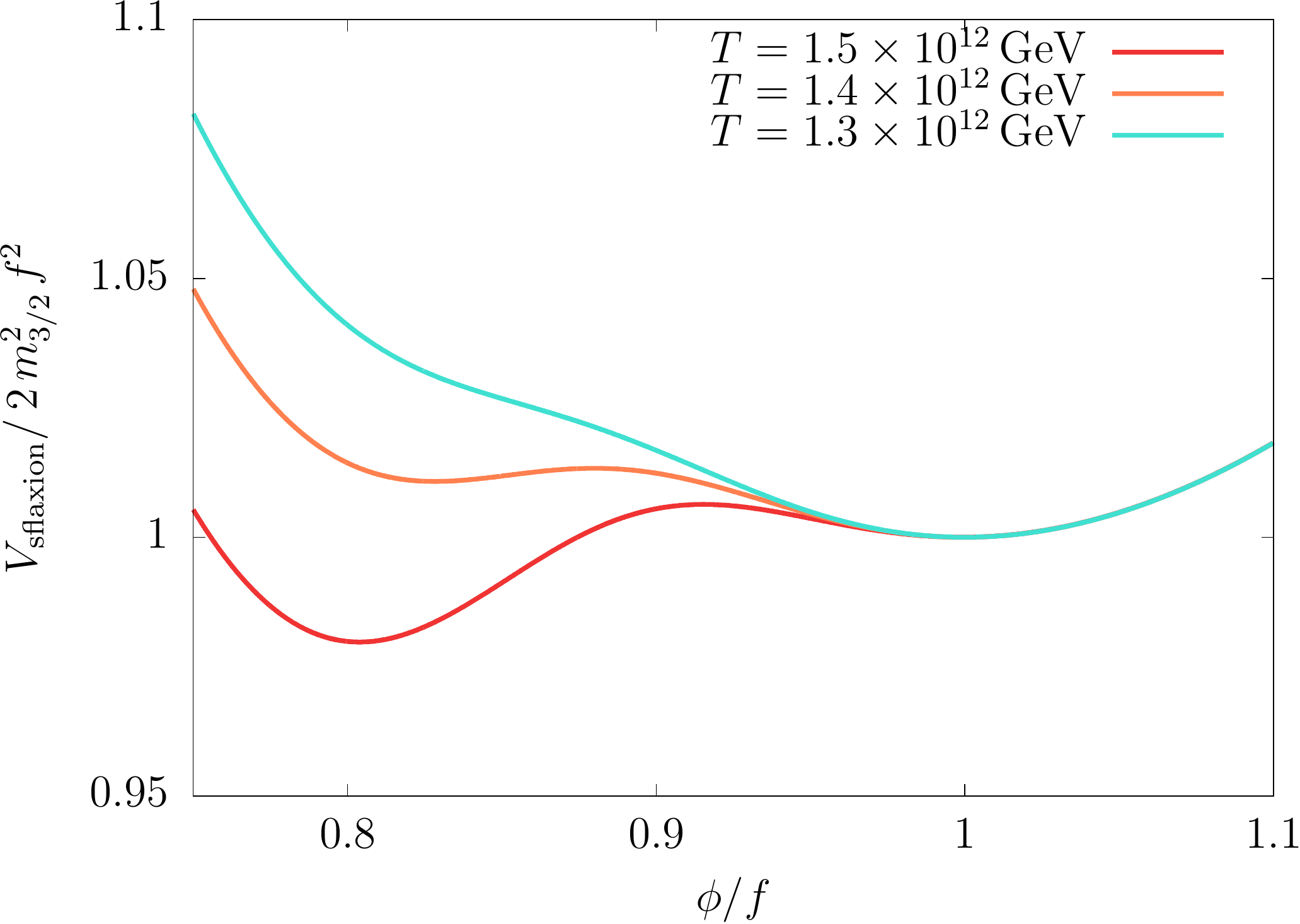}
		\end{center}
	\end{minipage}
	\begin{minipage}{0.5\hsize}
		\begin{center}
		\includegraphics[width=8cm]{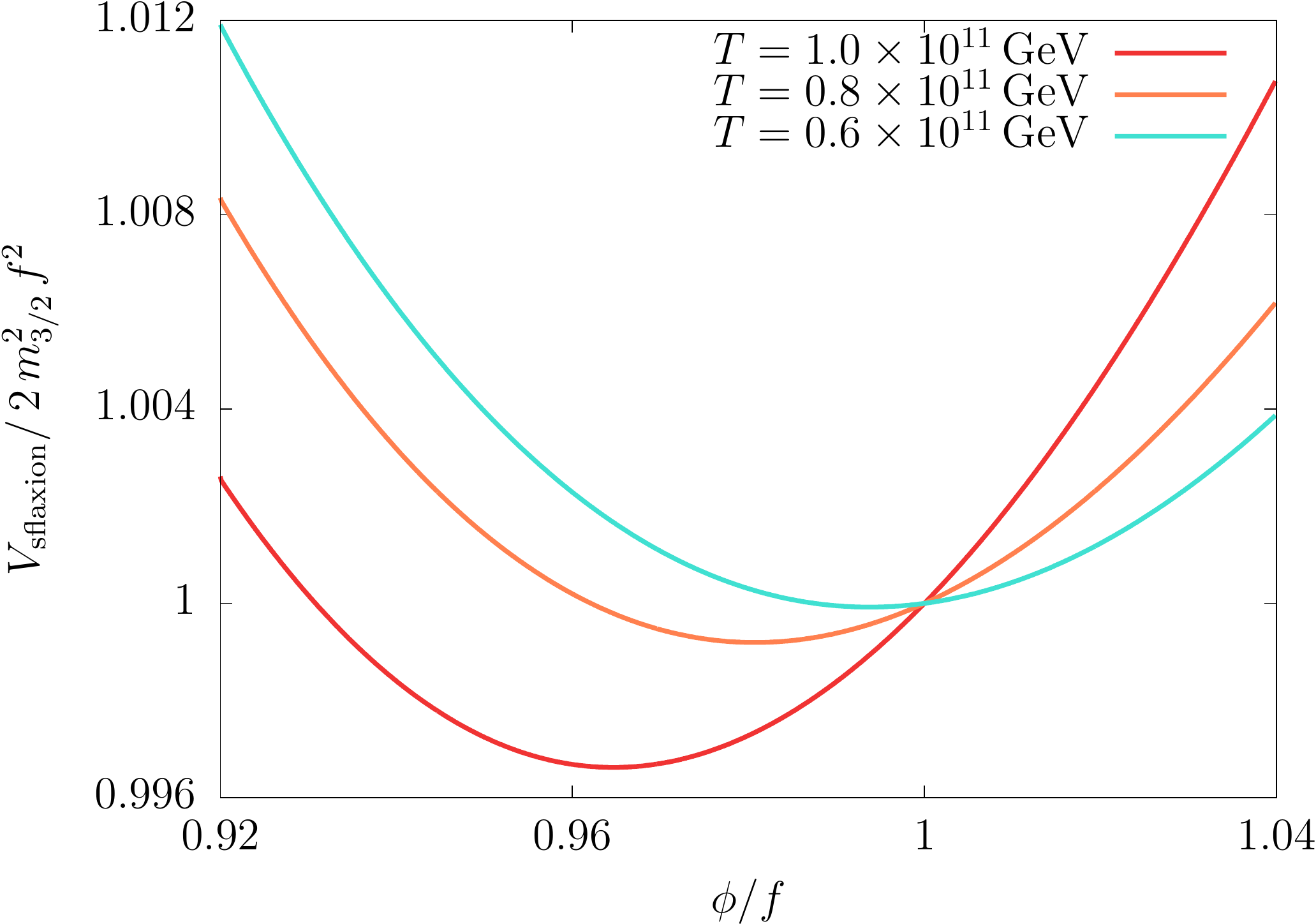}
		\end{center}
	\end{minipage}
	\caption{
	\small
	The changes of the sflaxion potential with respect to temperature. The cutoff scale is set to be $M=10^{19}\,\mathrm{GeV}$ and $10^{17}\, \text{GeV}$ in the left and right figure, respectively. Other parameters are the same: $n=9$, $y=1$, and $m_{3/2}=10^3\, \text{TeV}$. Here we take $V_\text{sflaxion}$ as $m_{3/2}^2 \phi^2+m_{3/2}^2 f^4/\phi^2 + V_T(M_\text{eff}(\phi),T)-V_T(M_\text{eff}(f),T)$. In the left figure, the trapping minimum disappears as the temperature falls, which induces the sflaxion oscillation. On the other hand, in the right figure, the trapping minimum and the true minimum are indistinguishable, which means no sflaxion oscillation. $(y\epsilon^n)/\sqrt{m_{3/2}/M}\simeq 6$ and 0.6 in the left and right figure, respectively (see Eq.~\eqref{no-oscillation}).}
	\label{fig:sflaxion_potential_shape}
\end{figure}

Now we study the sflaxion potential 
in order to estimate $\rho_{\sigma,i}/s_i$ and hence $\gamma$.
For the purpose of illustration, we consider a simple case where only the lightest right-handed neutrino $N_1$, with its exponent $n$ and its $\mathcal{O}(1)$ coupling $y$, is involved.\footnote{Note that $M$, $n$ and $y$ are related so as to give an appropriate mass for $N_1$. For example, thermal leptogenesis favors $m_{N_1} = y\epsilon^nM\gtrsim 10^{11}\, \text{GeV}$ (see Eq.\eqref{baryon_asymmetry_susy}).}
We assume that $T_R$ is comparable to or larger than $m_{N_1}$ for the thermalization of $N_1$. 
As long as $T_R$ is not so high as to thermalize $N_{2(3)}$, the effect of $N_{2(3)}$ can be neglected. 
Moreover, the effect of the $\mu$-term is subdominant due to its large exponent
for the temperature of our interest.
We also neglect thermal effect coming from the stabilizer field $S$, which is justified unless $m_S \sim \lambda f \sim m_{N_1}$.
Then, by applying the relation $\phi\overline\phi=f^2$, the sflaxion potential for $T \gtrsim M_\text{eff}$ is given by
\begin{align}
	V_\text{sflaxion}(\phi,T) \simeq m_{3/2}^2 \phi^2+m_{3/2}^2 \frac{f^4}{\phi^2}+\frac{y^2}{8}M^2T^2\left(\frac{\phi}{M}\right)^{2n},
	\label{potential_2.6}
\end{align}
where we have assumed that the same soft mass $m_{3/2}$ is generated for $\phi$ and $\overline\phi$. 
In general, the sflaxion may be trapped at a (local) minimum other than the true minimum at high temperature, which we call trapping minimum $\phi_t(T)$~\cite{Moroi:2012vu}. 
After the thermal effects become insignificant, the temporal minimum of the sflaxion potential approaches to the true minimum.
If the transition is not smooth, the sflaxion coherent oscillation is induced. 
See the left figure in Fig.~\ref{fig:sflaxion_potential_shape}.
Now let us investigate the sflaxion evolution by dividing it into two cases: the sflaxion oscillation is induced or not.

\begin{itemize}

\item {\bf Case without sflaxion oscillation:}
If the trapping minimum smoothly merges to the true minimum, the sflaxion oscillation is not induced and there is no dilution. The temperature $T_i$, when the thermal effect of $N_1$ becomes negligible, can be estimated by $M_\text{eff}(\phi_t(T_i))=T_i$. We can suppose that the trapping minimum and the true minimum cannot be distinguished if the soft mass overcomes the thermal effect at $\phi_t(T_i)\, (\simeq f)$ and $T_i$. Thus, the condition that the sflaxion oscillation is not induced is roughly given by
\begin{align}
	2m_{3/2}^2f^2 \gtrsim \frac{y^4}{8}M^4\left(\frac{f}{M}\right)^{4n},
\end{align}
hence
\begin{align}
	y\epsilon^{n} \lesssim \sqrt{\frac{m_{3/2}}{M}}.
	\label{no-oscillation}
\end{align}
This condition is checked by studying the temperature dependence of
the shape of the potential \eqref{potential_2.6} with its thermal contribution part replaced by the exact expression as the first one in \eqref{V_T}. In general,
it can be satisfied for large $n$.  For example, if we choose $m_{3/2}=10^3\, \text{TeV}$,
$n=9$, $y=1$, and $M=10^{17}\, \text{GeV}$, which leads to $m_{N_1}\simeq 2\times 10^{11}\, \text{GeV}$,
the sflaxion oscillation is not induced as deduced from the right panel of Fig.~\ref{fig:sflaxion_potential_shape}.
(In this case, for $\mu\simeq m_{3/2}$, we also obtain $n^\mu=17$, $N_\text{DW}=28$, and $f_a\simeq
2\times 10^{15}\, \text{GeV}$.)  Thus there is no late-time
entropy production and a sufficient amount of baryon asymmetry can be
produced through thermal leptogenesis.

\item {\bf Case with sflaxion oscillation:} If the condition
  \eqref{no-oscillation} is not satisfied, the sflaxion oscillation is
  induced.  In this case, because $\phi_t(T_i)$ is not close to $f$,
  we can neglect the first term in \eqref{potential_2.6} and
  obtain\footnote{Because $\overline\phi$ can become larger than the
    cutoff scale $M$ for this value of $\phi$, there may be some
    corrections in general. However, such effects are irrelevant at
    least for our choice of \eqref{eq:Wflaxion} and
    \eqref{eq:Kflaxion}.}
\begin{align}
	\phi_t(T) \simeq M\left(\frac{8\epsilon^4}{ny^2}\frac{m_{3/2}^2}{T^2}\right)^{1/(2n+2)},
	\label{trapping_min}
\end{align}
from $\partial_\phi V_\text{sflaxion}(\phi_t(T),T)=0$.
Then we find $T_i$ that satisfies $M_\text{eff}(\phi_t(T_i))=T_i$ as
\begin{align}
	T_i \simeq	 y^{1/(2n+1)}M\left(\frac{8\epsilon^4}{n}\frac{m_{3/2}^2}{M^2}\right)^{n/(4n+2)}.
	\label{critical_temp}
\end{align}
Combining \eqref{dilution_factor}, \eqref{decay_temperature}, \eqref{trapping_min}, and \eqref{critical_temp} with
\begin{align}
	\frac{\rho_{\sigma,i}}{s_i} \simeq \frac{m_{3/2}^2\frac{f^4}{\phi_i^2}}{\frac{2\pi^2}{45}g_*T_i^3} = \frac{45}{2\pi^2 g_*}\frac{m_{3/2}^2 f^4}{T_i^3 \phi_i^2},
\end{align}
we get, for the case of $\gamma<1$,
\begin{align}
	\gamma &\simeq 1.6\times 10^{11} \left(\frac{8\cdot 0.23^{2n+4}}{n\cdot 10^{12}}\right)^{\frac{3n+2}{4n+2}} y^{\frac{3n+1}{2n+1}} \left(\frac{n^\mu}{10}\right) \left(\frac{m_{3/2}}{10^3\,\text{TeV}}\right)^{\frac{4n+3}{4n+2}} \left(\frac{10^{12}\,\text{GeV}}{m_{N_1}}\right)^{\frac{3n+2}{2n+1}} \label{dilution_factor_simplified} \\
	&\simeq 
	\begin{cases}
		6\times 10^{-2} \cdot y^ {4/3}\left(\frac{n^\mu}{10}\right)\left(\frac{m_{3/2}}{10^3\,\text{TeV}}\right)^{7/6} \left(\frac{10^{12}\,\text{GeV}}{m_{N_1}}\right)^{5/3}~~~\text{for $n=1$} \\
		1\times 10^{-3} \cdot y^{10/7}\left(\frac{n^\mu}{10}\right)\left(\frac{m_{3/2}}{10^3\,\text{TeV}}\right)^{15/14} \left(\frac{10^{12}\,\text{GeV}}{m_{N_1}}\right)^{11/7}~~~\text{for $n=3$}
	\end{cases},
\end{align}
where we have set $\mu=m_\sigma=m_{3/2}$ for simplicity. Unless $m_{3/2}$ or $y$ is extremely large, $\gamma$ is typically very small, which is preferred for diluting dangerous relics but undesirable for thermal leptogenesis. 
By using  \eqref{baryon_asymmetry_susy} and \eqref{dilution_factor_simplified}, we can investigate parameters for which thermal leptogenesis is compatible with dilution. 
For example, if we choose $m_{3/2}=4\times 10^{3}\, \text{TeV}$, $n=1$, $y=1$, and $M=2\times 10^{12}\, \text{GeV}$, which leads to $m_{N_1}\simeq 5\times 10^{11}\, \text{GeV}$, 
the proper amount of baryon asymmetry can be reproduced.
(In this case, for $\mu\simeq m_{3/2}$, we obtain $n^\mu=9$, $N_\text{DW}=4$, and $f_a\simeq 2\times 10^{11}\, \text{GeV}$.) However, in general, as $n$ increases, it becomes more difficult to find successful parameters.
\end{itemize}

As seen in the above two cases, there exist parameters for successful thermal leptogenesis even in the presence of the sflaxion, although they are subject to severe constraints. If $T_R$ is not so high as to thermalize $N_1$, or if $n^N_{\alpha\beta}=0$, the similar discussion with $N_1$ replaced by the Higgs fields may be applied; in this case, the sflaxion oscillation may not be induced due to the large exponent. In general, the more fields get involved in the sflaxion evolution, the more complex the situation becomes and the above simplified picture can be significantly changed.

\section{Inflation in SUSY flaxion model}
\label{sec:inflation}

\subsection{Multi-field generalization of attractor inflation model}
\label{sec:alpha-attractor}

In this section we discuss possible implementation of inflation in the SUSY flaxion model.
Let us take the K\"ahler potential and superpotential as (we use the Planck unit $M_P=1$ in this section)
\begin{align}
	&K = -\Lambda^2\log\left[ \frac{(1-|\Phi|^2/\Lambda^2)(1-|\overline\Phi|^2/\Lambda^2)}{(1-\Phi\overline\Phi/\Lambda^2)(1-\Phi^\dagger\overline\Phi^\dagger/\Lambda^2)} \right] + |S|^2 - c|S|^4,
	\label{K_original} \\
	&W = \kappa S (\Phi \overline\Phi - \mathcal F^2) + W_0 + W_{\text{MSSM}+N},   \label{W_original}
\end{align}
where $\Lambda$ is some mass scale, which will turn out to be close to the PQ scale later, 
and $c$, $W_0$, $\kappa$ and $\mathcal F$ are assumed to be real and positive.\footnote{
	In general, $\mathcal F$ cannot be taken to be real in the basis in which the coefficient of 
	$\Phi\overline\Phi$ term in the K\"ahler potential is taken to be real.
	In the following we assume $\mathcal F$  is real for simplicity, although the results do not much depend on this assumption.
}
\footnote{
	The suppression scale of $\Phi$ and $\overline\Phi$ inside the logarithm in \eqref{K_original} can be different from the overall coefficient of the first term in the K\"ahler potential. Changing the suppression scale effectively changes the mass scale $M$ appearing as the coefficient of the bilinear terms in $W_{\text{MSSM}+N}$.
}
The relation between the PQ fields $(\Phi, \overline\Phi)$ and $(\phi,\overline\phi)$ in the previous section will be explained
at the end of this subsection.
This may be a natural multi-field generalization of the $\alpha$-attractor model in supergravity~\cite{Kallosh:2013hoa,Kallosh:2013daa,Kallosh:2013yoa,Carrasco:2015uma,Kallosh:2016gqp}. (See also Ref.~\cite{Yamada:2018nsk} for a related work.)
Although there may be some other inflation models that can be consistently implemented in the SUSY flaxion model,
we focus on this particular setup in this section for concreteness.

Let us see that this model has a flat potential for large PQ field values as required for inflation.
The kinetic term and the scalar potential read
\begin{align}
  {\cal L}_{\text{kin}} &=
  \frac{1}{(1-|\Phi|^2/\Lambda^2)^2} |\partial \Phi|^2
  +
  \frac{1}{(1-|\overline\Phi|^2/\Lambda^2)^2} |\partial \overline\Phi|^2,
\end{align}
and
\begin{align}
  V = h(\Phi,\overline\Phi)^{\Lambda^2} \kappa^2 
  \left|\Phi\overline\Phi-\mathcal F^2\right|^2,
  \label{V}
\end{align}
with 
\begin{align}
  h(\Phi,\overline\Phi) \equiv
  \frac{(1-\Phi\overline\Phi/\Lambda^2)(1-\Phi^\dagger\overline\Phi^\dagger/\Lambda^2)}
  {(1-|\Phi|^2/\Lambda^2)(1-|\overline\Phi|^2/\Lambda^2)}.
\end{align}
Note that here we have taken $S=0$, which will be justified later.

\begin{figure}[t]
		\begin{center}
		\includegraphics[width=8cm]{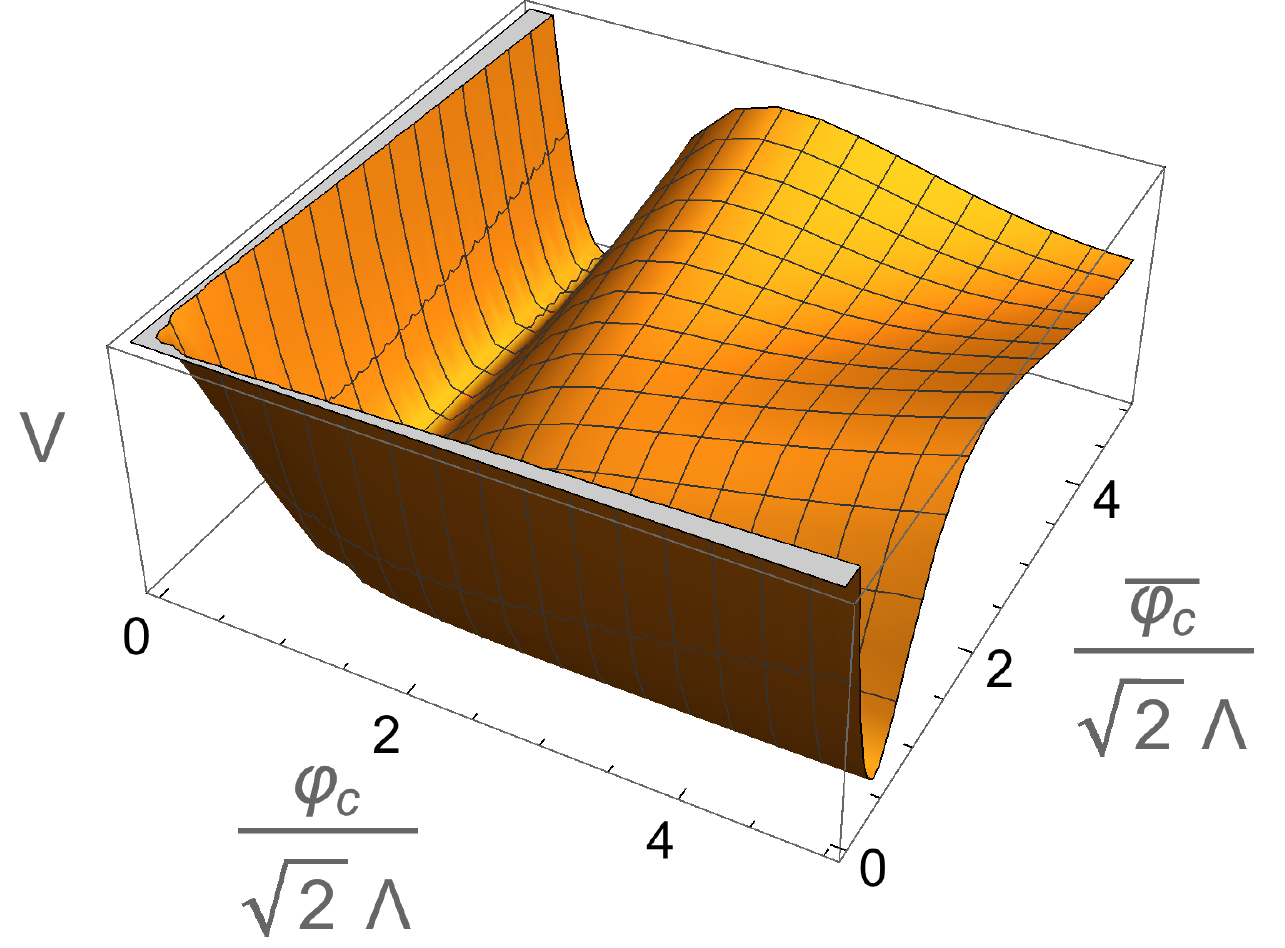}
		\includegraphics[width=8cm]{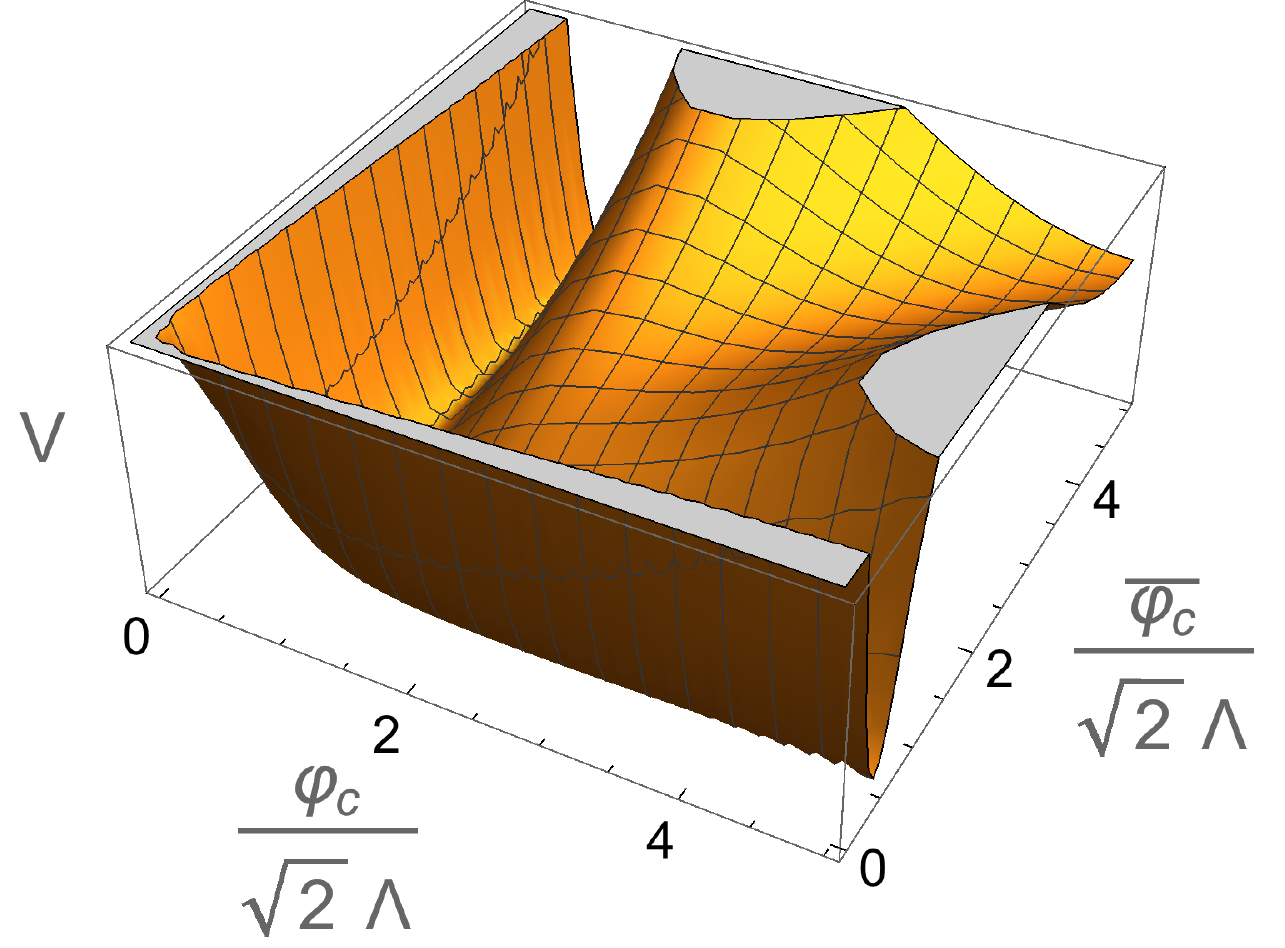}
		\end{center}
	\caption{
	The inflaton potential in terms of the canonical fields $\varphi_c$ and $\overline\varphi_c$.
	We have taken $\mathcal F/\Lambda = 0.75$ and $\Lambda^2=0.1$ (left) and $\Lambda^2=0.2$ (right).
	}
	\label{fig:pot}
\end{figure}

In order to see the inflaton dynamics, let us parametrize the fields as
\begin{align}
	\Phi = \varphi e^{i\theta},~~~~~~\overline\Phi=\overline\varphi e^{i\overline\theta}.
\end{align}
For a while, let us assume that phase is stabilized as $\theta + \overline\theta=0$, which will be justified later.
The orthogonal field is the massless mode corresponding to the flaxion.
Then focusing on the real part of the PQ fields, we obtain the canonically normalized fields as
\begin{align}
	\varphi=  \Lambda\tanh\left(\frac{\varphi_c}{\sqrt{2}\Lambda}\right),~~~~~ \overline\varphi=  \Lambda\tanh\left(\frac{\overline\varphi_c}{\sqrt{2}\Lambda}\right).
	\label{phi_canonical}
\end{align}
Using them, we can define the canonical inflaton field $\chi$ and sflaxion $\sigma$ as
\begin{align}
	\chi = \frac{1}{\sqrt 2}(\varphi_c + \overline\varphi_c),~~~~~~~~
	\sigma= \frac{1}{\sqrt 2}(\varphi_c - \overline\varphi_c).
\end{align}
Moreover, let us suppose that the sflaxion field is stabilized at $\sigma=0$ during inflation,
which will also be justified later, and consider the field trajectory along $\varphi= \overline\varphi$.
Noting that $h(\Phi,\overline\Phi)=1$ along this direction, the scalar potential is given by 
\begin{align}
	V = \kappa^2\left[\Lambda^2\tanh^2\left(\frac{\chi}{2\Lambda}\right) -\mathcal F^2\right]^2,
\end{align}
Thus it has a flat potential for $\chi \gg 2\Lambda$ that is suitable for inflation,
and it is the same as the inflation model of \cite{Ema:2016ops}.
The shape of the scalar potential in terms of the canonical fields $\varphi_c$ and $\overline\varphi_c$ is shown in Fig.~\ref{fig:pot},
where it is seen that the potential becomes flat along the inflaton direction $\varphi_c = \overline\varphi_c$.
We impose $1/\sqrt 2 < \mathcal F/\Lambda < 1$ so that the inflaton potential energy is lower than the potential energy at the origin
and no PQ symmetry restoration occurs after inflation, meaning that $\mathcal F$ is close to $\Lambda$.
The canonical inflaton field value at the $e$-folding number $N_e$ before the end of inflation is given by
\begin{align}
  \chi (N_e) \simeq \Lambda \ln \left(\frac{8N_e}{\Lambda^2 \Delta}\right),
\end{align}
where
\begin{align}
  \Delta \equiv 1- \mathcal F^2/\Lambda^2,
\end{align}
which is in the range of $0<\Delta < 1/2$.  The field value in the
original basis is given by
\begin{align}
	\varphi (N_e) \simeq \Lambda\left( 1-2\exp\left( -\frac{\chi}{\Lambda}\right) \right)
	\simeq \Lambda\left( 1-\frac{\Lambda^2 \Delta}{4N_e}.  \label{Phi_N}\right)
\end{align}
The scalar spectral index and the tensor-to-scalar ratio in this model are calculated as
\begin{align}
	n_s \simeq 1 - \frac{2}{N_e},~~~~~~~~~~r\simeq \frac{8\Lambda^2}{N_e^2}.
\end{align}
They fall into the best-fit values of the Planck observation for $N_e=50$--$60$.
The power spectrum of the curvature perturbation is evaluated as
\begin{align}
	\mathcal P_\zeta \simeq \frac{N_e^2}{12\pi^2}\kappa^2\Lambda^2\Delta^2.  \label{P_zeta}
\end{align}
For reproducing the observed magnitude of the density perturbation of the universe $\mathcal P_\zeta \simeq 2.2\times 10^{-9}$~\cite{Ade:2015xua},
we need
\begin{align}
	\Lambda \simeq 2.5\times 10^{13}\,{\rm GeV} \left(\frac{1}{\kappa\Delta}\right)\left( \frac{50}{N_e} \right).
	\label{kappa}
\end{align}

Some comments are in order.
In our inflation model in which PQ fields are identified as the inflaton, 
the PQ symmetry is already spontaneously broken during inflation and never restored thereafter.
Thus it does not suffer from an axionic DW problem.\footnote{
	It is possible to arrange the U(1)$_{\rm F}$ charges so that the DW number is equal to unity.
	In this case we do not necessarily have a DW problem. 
}
Interestingly, the flaxion isocurvature perturbation is also significantly suppressed and easily satisfies the observational constraint
due to the combined effect of effectively large PQ scale and enhanced flaxion kinetic term during inflation.
We refer to the previous work~\cite{Ema:2016ops} in this respect
since the story is the same as the non-SUSY flaxion model.
The resulting power spectrum of the dark matter isocurvature perturbation, divided by the adiabatic perturbation, is given by\footnote{Note that the normalization of $\Lambda$ is different from Ref.~\cite{Ema:2016ops} by a factor of $\sqrt{2}$. The formula of ${\cal P}_S/{\cal P}_\zeta$ in Ref.~\cite{Ema:2016ops} is also missing a factor of 4.
Combining them, Eq.~\eqref{eq:isocurvature} differs by a factor of 16 compared with Ref.~\cite{Ema:2016ops}.
}
\begin{align}
\frac{{\cal P}_S}{{\cal P}_\zeta}\simeq R_a^2\frac{N_{\rm DW}^2\Delta^2 \Lambda^4}{4N_e^4 \Theta_i^2},
\label{eq:isocurvature}
\end{align}
where $R_a$ is the fraction of the flaxion dark matter in the total dark matter abundance
and $\Theta_i$ denotes the initial misalignment angle of the flaxion $(0<\Theta_i < 2\pi)$.
Observationally this ratio is constrained as ${\cal P}_S/{\cal P}_\zeta \lesssim 0.04$~\cite{Ade:2015lrj},
which is easily satisfied for a reasonable choice of $\Lambda$, as we shall see below.

Now let us explain the relation between the PQ fields $(\Phi, \overline\Phi)$ and $(\phi,\overline\phi)$ in the previous section.
The superpotential term $W_{\text{MSSM}+N}$ in (\ref{W_original}) has the same form as (\ref{eq:W_MSSM})
after replacing $\phi$ in (\ref{eq:W_MSSM}) with $\Phi$.
Thus we have $\epsilon = \langle\Phi\rangle / M = \mathcal F / M \simeq 0.23$.
Since $1/\sqrt 2 < \mathcal F/\Lambda < 1$ for successful inflation, the cutoff scale $M$ should also be close to (or a bit larger than) $\Lambda$.
All the flavor structure is reproduced after $\Phi$ gets a VEV as explained in the previous section.
However, one should note that the VEV of $\Phi$ itself does not correspond to the physical PQ scale, 
since $\Phi$ is not canonically normalized.
As far as phenomena around the potential minimum $\langle\Phi\rangle=\langle\overline\Phi\rangle=\mathcal F$ are concerned,
we can simply define the canonical fields as
\begin{align}
	\phi-\langle\phi\rangle = \frac{1}{\Delta} (\Phi-\mathcal{F}),~~~~~~~\overline\phi-\langle\overline\phi\rangle = \frac{1}{\Delta} (\overline\Phi-\mathcal{F}).
\end{align}
Thus from (\ref{eq:fa_phi_M_1}), the PQ scale is given by
\begin{align}
	f_a = \frac{2 \mathcal F}{N_{\rm DW}} \frac{1}{\Delta} =  \frac{2 \epsilon M}{N_{\rm DW}} \frac{1}{\Delta}.
	\label{fa_alpha}
\end{align}
Recalling that $\mathcal F$ (and also $M$) is close to $\Lambda$ and $\Delta$ is about $1/2$,
the mass scale $\Lambda$ effectively determines the PQ scale.

\subsection{Stability of inflationary path}

Now we will see that $h(\Phi,\overline\Phi)=1$ is dynamically realized along the inflationary trajectory.
Noting that $h$ is rewritten as
\begin{align}
	h (\Phi,\overline\Phi) 
	= \frac{1-2(\varphi\overline\varphi/\Lambda^2)\cos(\theta+\overline\theta) + \varphi^2\overline\varphi^2/\Lambda^4}{(1-\varphi^2/\Lambda^2)(1-\overline\varphi^2/\Lambda^2)},
\end{align}
it is seen that $h$ is minimized for $\theta+ \overline\theta = 0$.
By expanding the scalar potential up to the quadratic term in $\theta
+ \overline\theta$ around the inflationary trajectory
$\varphi=\overline\varphi$, we find a mass term like
\begin{align}
	V \supset \left[ \mathcal F^2 \left(1-\frac{\varphi^2}{\Lambda^2}\right)^2 + (\varphi^2-\mathcal F^2)^2 \right] \kappa^2\left(\frac{\theta_c + \overline\theta_c}{\sqrt{2}}\right)^2,
\end{align}
where we have defined canonical phase fields as $\theta_c \equiv \sqrt{2} \varphi \theta/ (1-\varphi^2/\Lambda^2)$ and 
$\overline\theta_c \equiv \sqrt{2} \overline\varphi \overline\theta/ (1-\overline\varphi^2/\Lambda^2)$.
During inflation, the first term is negligible according to (\ref{Phi_N}) and the second term gives
$m^2_{\theta+\overline\theta} = 6H_{\rm inf}^2$,
where $H_{\rm inf}$ denotes the Hubble scale during inflation and satisfies $3H_{\rm inf}^2 = \kappa^2 (\varphi\overline\varphi - \mathcal F^2)^2$.
On the other hand, well after inflation the first term is the dominant one and it gives
$m^2_{\theta+\overline\theta} = 2\kappa^2 \mathcal F^2 \Delta^2$.
Thus we can safely take $\theta+ \overline\theta = 0$ during as well as after inflation.
The orthogonal one, $\theta- \overline\theta$, corresponds to the flaxion.

Since we can take $\theta+ \overline\theta = 0$, the function $h$ is simplified as
\begin{align}
	h (\varphi,\overline\varphi) 
	= \frac{(1-\varphi\overline\varphi/\Lambda^2)^2}{(1-\varphi^2/\Lambda^2)(1-\overline\varphi^2/\Lambda^2)},
\end{align}
during inflation. Apparently, $h=1$ for the inflationary trajectory $\varphi=\overline\varphi$
and we want to see the fluctuation about it.
The function $h$ is expressed in terms of the canonical fields as
\begin{align}
	h (\varphi,\overline\varphi)  
	= \frac{\left[1-\tanh\left(\frac{\varphi_c}{\sqrt{2}\Lambda}\right)\tanh\left(\frac{\overline\varphi_c}{\sqrt{2}\Lambda}\right) \right]^2}
	{\left[1-\tanh^2\left(\frac{\varphi_c}{\sqrt{2}\Lambda}\right)\right]\left[1-\tanh^2\left(\frac{\overline\varphi_c}{\sqrt{2}\Lambda}\right)\right]}
	= \cosh^2\left(\frac{\sigma}{\Lambda}\right)
\end{align}
It depends only on the sflaxion direction, which is orthogonal to the inflaton direction,
and has a positive curvature along the sflaxion direction.
Expanding the scalar potential up to the quadratic term in $\sigma$, the sflaxion mass term is found to be
\begin{align}
	V \supset \left[ (\varphi^2 - \mathcal F^2) - \frac{1- \varphi^4/\Lambda^4}{2}\right] (\varphi^2-\mathcal F^2) \kappa^2 \sigma^2.
\end{align}
During inflation, the second term in the square bracket can be neglected according to (\ref{Phi_N}) and the sflaxion mass is evaluated as $m_{\sigma}^2 = 6H_{\rm inf}^2$. Thus it is stabilized at $\sigma=0$, 
justifying the assumption of the stability of the inflationary trajectory $\varphi=\overline\varphi$.
After inflation ends, it becomes (temporary) tachyonic as seen in Fig.~\ref{fig:pot}.
Note that this expression shows vanishing sflaxion mass around the potential minimum $\varphi = \mathcal F$,
reflecting the fact that the sflaxion is a scalar partner of the flaxion,
although actually it gets the SUSY breaking mass after properly taking the SUSY breaking effect into account.

Finally, we also comment on the validity of taking $S = 0$ for calculating the inflaton potential.
It is known that the potential minimum deviates from $S=0$ during and after inflation
due to the constant term $W_0$ in (\ref{W_original}) which might induce a linear term in the scalar potential~\cite{Dvali:1997uq,Buchmuller:2000zm,Nakayama:2010xf}.
During the inflation, the induced linear term potential is $V\sim {\cal O}(m_{3/2} H_{\text{inf}} S)$, 
whereas the mass term is given by $V\sim  {\cal O}(cH_{\text{inf}}^2  |S|^2)$. Here, a nonzero $|S|^4$ term with $c>1/2$ is necessary to have a positive mass squared for $S$ during the inflation. 
Thus we have $S\sim m_{3/2}/cH_{\text{inf}} \sim m_{3/2} / [c\kappa(\varphi^2-\mathcal F^2)]$ during inflation.
It slightly modifies the inflaton potential, but we have checked that its effect on the inflaton dynamics is negligible as far as $m_{3/2}\ll cH_{\rm inf}$ and practically the potential is well approximated by (\ref{V}).
On the other hand, at the potential minimum, the linear term of the $S$ field vanishes because of the particular form of the K\"ahler potential (\ref{K_original}).
It is explicitly seen by noting that the first derivatives of the K\"ahler potential read
\begin{align}
	K_\Phi = \frac{\Phi^\dagger-\overline\Phi}{(1-|\Phi|^2/\Lambda^2)(1-\Phi\overline\Phi/\Lambda^2)},~~~~~~~~
	K_{\overline\Phi} = \frac{\overline\Phi^\dagger-\Phi}{(1-|\overline\Phi|^2/\Lambda^2)(1-\Phi\overline\Phi/\Lambda^2)},  \label{K_Phi}
\end{align}
which vanish along the inflationary trajectory and also at the potential minimum $\Phi = \overline\Phi = \mathcal F$.
As a result, the linear term in the scalar potential vanishes at the potential minimum.\footnote{
	To be precise, $K_\Phi$ and $K_{\overline \Phi}$ vanish at the potential minimum only if $\mathcal F$ is real.
	If $\mathcal F$ is not real, $S$ obtains a VEV of $\sim m_{3/2}$ 
	at the potential minimum and it gives the axino mass of $\sim m_{3/2}$.
}

\subsection{Reheating}

Now let us consider the reheating after inflation.
In our model, the inflaton may most efficiently decay into RH (s)neutrinos and/or (s)flaxions.
As for the inflaton decay into RH (s)neutrinos, the relevant term in the superpotential (\ref{eq:W_MSSM}) in our model basis (\ref{K_original}) is
\begin{align}
	W = \frac{1}{2}y^N_{\alpha\beta}\left( \frac{\Phi}{M} \right)^{n^N_{\alpha\beta}}M N_\alpha N_\beta.
\end{align}
After properly taking account of the canonical normalization of $\Phi$, the partial decay rate of the inflaton into the RH neutrinos is evaluated as\footnote{The decay rate into RH sneutrinos, $\Gamma(\chi\to \tilde{N}\tilde{N})$, is suppressed by the RH sneutrino masses 
compared with that into RH neutrinos.
Although the supergravity effect induces the (maximal) mixing of $\chi$ and $S$ which would result in the inflaton decay rate into RH sneutrinos comparable to that into RH neutrinos~\cite{Kawasaki:2006hm}, it is not important in our case since the decay is much faster than the mixing time scale.}
\begin{align}
	\Gamma(\chi\to NN) \simeq \sum_{\alpha\beta} \frac{(n^N_{\alpha\beta} y^N_{\alpha\beta}\epsilon^{n^N_{\alpha\beta}-1})^2}{64\pi} \Delta^2m_\chi,
\end{align}
if kinematically allowed. Here, the phase space factor is omitted for simplicity.
Note that the inflaton mass around the potential minimum, $m_\chi$, is almost fixed by the condition (\ref{kappa}):
\begin{align}
	m_\chi = \sqrt 2 \kappa \mathcal F \Delta \simeq 2.8\times 10^{13}\,{\rm GeV}\, \left(\frac{\mathcal F/\Lambda}{0.8}\right) \left( \frac{50}{N_e} \right).
\end{align}
However, this inflaton decay mode is prohibited if either $n_{\alpha\beta}^N=0$ or the inflaton is lighter than the RH neutrinos.
Actually, in the next subsection we will see that $n_{\alpha\beta}^N=0$ is favored after taking account of the sflaxion dynamics.
In such a case, the inflaton dominantly decays into a pair of flaxions or sflaxions.
The inflaton decay rates into the flaxions and sflaxions are given by\footnote{
	To be precise, one should take account of the effects of 
	(tachyonic) preheating~\cite{Dolgov:1989us,Traschen:1990sw,Shtanov:1994ce,Kofman:1994rk,Kofman:1997yn,Felder:2000hj,Dufaux:2006ee}
	for the (s)flaxion production. 
	However, the following phenomenological consequence that (s)flaxions are efficiently produced and they are soon thermalized is not affected.
}
\begin{align}
	\Gamma(\chi\to aa)= 
	\Gamma(\chi\to \sigma\sigma)= 
	\frac{1}{128\pi} \frac{m_\chi^3 \Delta^2}{\mathcal F^2}. 
	\label{Gamma_aa}
\end{align}
We assume that the produced flaxions and sflaxions are soon thermalized through the processes like 
$aa/\sigma\sigma\to t \bar c H_u$, which is justified below.
Since the inflaton decay rate is typically much larger than the Hubble scale after inflation,
the reheating is completed almost instantaneously.
Thus the reheating temperature is expected to be comparable to the scale of PQ breaking.
Using (\ref{kappa}), the reheating temperature is estimated as
\begin{align}
	T_{R} \simeq \left( \frac{30}{\pi^2 g_*} \right)^{1/4}\Lambda\sqrt{\kappa\Delta}
	\simeq 8.4\times 10^{12}\,{\rm GeV}\left( \frac{1}{\kappa\Delta} \right)^{1/2}\left(\frac{50}{N_e} \right).
	\label{reheating_temperature}
\end{align}
Thus the reheating temperature may be more or less comparable to the PQ scale.\footnote{
	Note however that the PQ symmetry is never restored because of the requirement $1/\sqrt{2} < \mathcal F/\Lambda < 1$.
	One also needs not worry about possible DW formation due to the resonant enhancement of the flaxion fluctuation~\cite{Ema:2017krp,Kawasaki:2017kkr}.
}

Now let us show that the produced flaxions and sflaxions can be thermalized 
in the case that the inflaton main decay modes are $\chi\to aa$ and $\chi\to \sigma\sigma$.
In our charge assignment in Table~\ref{tab:charges}, the processes $aa/\sigma\sigma\rightarrow Q_3 \overline U_2 H_u$ may be dominant. 
The scattering rate is estimated as
\begin{align}
	\Gamma_{\rm scat}(aa/\sigma\sigma\rightarrow Q_3\overline U_2 H_u) \simeq n_{a/\sigma}\frac{3\epsilon^2}{1024\pi^3}(y^u_{32})^2 A\left(\frac{\Delta^4}{(\mathcal F/\Lambda)^4}\right) \frac{m_\chi^2}{\Lambda^4},
\end{align}
where $n_a\simeq n_\sigma\simeq (3H_\text{inf}^2M_P^2)/m_\chi$ is the number density of the flaxions and sflaxions. $A=\langle s^2/(s-m_\chi^2)^2\rangle$ is a factor of order unity ($A>1$), where $s$
is the Mandelstam variable, and $\langle \cdot \rangle$ 
represents the average over the momentum distributions of the flaxions and sflaxions.\footnote{
	Several effects may broaden the momentum distributions of the flaxion/sflaxion around $|\vec p| = m_\chi/2$:
	preheating, Hubble expansion and self-scattering of flaxion/sflaxion.
}
For thermalization of the flaxion and sflaxion, the scattering rate must be larger than the Hubble scale after inflation. This requires $\Gamma_{\rm scat}\gtrsim H_\text{inf}$,
which results in, using (\ref{kappa}),
\begin{align}
(y^u_{32})^2 A\; (\kappa\Delta^2)^3 g(\Delta) 
\left(\frac{N_e}{50}\right)
\gtrsim 0.8
	\label{Lambda_upperbound}
\end{align}
where $g(\Delta)=\Delta (1-\Delta)^{-3/2}$ ($0<g\lesssim \sqrt{2}$).
This can be satisfied with couplings of order unity,  and it is compatible with the 
perturbativity bound on the inflaton coupling, $\kappa\Delta^2\lesssim 4\pi$. (Note that $\kappa\Delta^2$ is the coupling constant for the canonically normalized fields.)
From \eqref{kappa} and \eqref{reheating_temperature}, the order one coupling
$\kappa\Delta^2\simeq {\cal O}(1)$ implies that the $\Lambda$ and $T_R$ are more or less fixed at around $\Lambda\simeq {\cal O}(10^{13})~\text{GeV}$ 
and $T_R\simeq {\cal O}(10^{12})~\text{GeV}$.

Here we comment on the nonthermal gravitino production.
It is known that for inflation models with non-vanishing inflaton VEVs the inflaton may directly decay into a pair of gravitinos,
which often leads to the gravitino overproduction~\cite{Endo:2006zj,Nakamura:2006uc,Kawasaki:2006gs,Asaka:2006bv,Dine:2006ii,Endo:2006tf,Kawasaki:2006hm,Endo:2007ih,Endo:2007sz}.
Interestingly, however, our model defined in (\ref{K_original}) and (\ref{W_original}) may avoid the nonthermal gravitino production
due to the property $K_\Phi = K_{\overline\Phi} = 0$ from (\ref{K_Phi}).
Actually we can easily find that all $\left<G_i\right> \equiv \left<K_i + W_i/W\right> = 0$, with $i = \Phi, \overline\Phi, S$ collectively denoting 
the field derivative, are the solution of the vacuum except for the SUSY breaking field $Z$,
if there is no direct coupling between $Z$ and other fields.\footnote{
	Again this statement is true only if $\mathcal F$ is real. Otherwise, nonthermal gravitino production is inevitable
	but thermal production is much more efficient in our model.
}
Thus we can safely neglect the nonthermal contribution to the gravitino abundance,
while thermally produced ones give significant contribution due to the high reheating temperature.
As shown in Sec.~\ref{sec:gravitino}, the gravitino problem can be solved thanks to the existence of $R$-parity violating operator 
in our model anyway.

\subsection{Successful scenario after inflation}
\label{sec:alpha_inf_sflaxion}

In this subsection, we show that there exist viable parameters for a consistent cosmological scenario including the reheating, the baryogenesis, and the dark matter in the present model.
In the following discussion, we suppose that the baryon asymmetry is generated by thermal leptogenesis and the dominant dark matter component is coherent oscillation of the flaxion.

For a successful scenario, a particular attention has to be paid to the sflaxion evolution after the reheating.
First, we consider the case that there exist non-zero $n^N_{\alpha\beta}$, \textit{i.e.}, the sflaxion couples to RH (s)neutrinos. 
Assuming viable thermal leptogenesis, RH (s)neutrinos are likely to be thermalized and hence
the finite-temperature effect from the RH (s)neutrinos can significantly affect the sflaxion potential, as shown in Sec.~\ref{sec:oscillation}.
In our particular model defined by (\ref{K_original}) and (\ref{W_original}), such a thermal effect can be more problematic.
To see this, let us recall that the sflaxion direction is defined by $\varphi\overline\varphi = \mathcal{F}^2$ along which the scalar potential vanishes
in the SUSY limit, but it obtains a mass after taking the SUSY breaking effect into account.
The scalar potential for the sflaxion direction is given by 
\begin{align}
  V \supset 
  h^{\Lambda^2} \frac{2m_{3/2}^2 }{\Delta^2}
  \left( \varphi^2 + \frac{\mathcal{F}^4}{\varphi^2} \right),
  \label{soft_mass}
\end{align}
where we have used $\varphi\overline\varphi = \mathcal{F}^2$ and $W_0=m_{3/2}M_P^2$.\footnote{
	Note that the factor $h$ diverges at $\varphi = \mathcal{F}^2/\Lambda$ and $\varphi=\Lambda$; however, because the exponent $\Lambda^2$ is  small, it is effective only at the neighborhood of these points, while in the other region the correction is negligible. 
}
In terms of the canonical fields (\ref{phi_canonical}), the potential is very flat in the region away from 
$\varphi = \mathcal{F}$ due to the singular behavior of the kinetic term in the original basis.
Note that the height of the potential of this flat region is $\sim m_{3/2}^2 \Lambda^2$, which is much lower than the inflaton potential.
Since thermal effects tend to bring $\varphi \to 0$, the sflaxion can easily go into this extremely flat region after inflation.
Once this happens, the sflaxion remains there for a long time even after the thermal effect becomes negligible, which causes another inflation.
This results in a large scale density perturbation inconsistent with observation and also dilution of the baryon asymmetry. 
Actually the thermal effect from $N_\alpha$ on the sflaxion potential is inevitable for non-zero $n^N_{\alpha\beta}$, which may take the sflaxion into the unwanted flat region.\footnote{As opposed to $N_\alpha$, the thermal effect of $S$ (and the orthogonal fields in $\Phi$ and $\overline\Phi$) which couples to both $\varphi$ and $\bar{\varphi}$ tends to stabilize the sflaxion at $\varphi=\mathcal F$, if it is thermalized. If $m_{N_1}$ is comparable to (or a little bit larger than) $m_S$ ($\simeq m_\chi$), the sflaxion can be stabilized at around $\varphi=\mathcal F$ through the evolution. However, in this case, it is difficult to find viable parameters to make the reheating temperature high enough for thermal leptogenesis and realize thermalization after the reheating at the same time.}
Thus in the following discussion we focus on the case of $n_{\alpha\beta}^N = 0$.\footnote{
	In this case we need to impose lepton parity to forbid the dangerous lepton-number violating operators
	since $\ell$ is integer, as shown in Sec.~\ref{sec:Rparity}.
}
Although there exist other thermal effects on the sflaxion even in this case, we checked that the sflaxion is not taken into the flat region by them through the evolution.
As mentioned in the previous subsection, the inflaton dominantly decays into the (s)flaxion pair
but they are soon thermalized if the condition (\ref{Lambda_upperbound}) is satisfied.

Let us check that the flaxion dark matter and thermal leptogenesis are compatible with the present scenario. As discussed in the previous subsection, the thermalization of the flaxion/sflaxion requires $\Lambda\simeq {\cal O}(10^{13})~\text{GeV}$ 
and $T_R\simeq {\cal O}(10^{12})~\text{GeV}$.
First, we consider the condition that coherent oscillation of the flaxion is the dominant dark matter component. From \eqref{kappa} and \eqref{fa_alpha},
\begin{align}
f_a &=4\times 10^{12}\,\text{GeV}
\left(\frac{1}{\kappa\Delta^2}\right)
\left(\frac{\mathcal F/\Lambda}{0.8}\right)
\left(\frac{10}{N_\text{DW}}\right) 
\left(\frac{50}{N_e}\right).
\end{align}
Thus, from \eqref{flaxion_DM} with $\Omega_ah^2\simeq 0.12$, the flaxion dark matter can be realized by taking $\Theta_i\lesssim \mathcal O(0.1)$.
Next, we consider the condition for thermal leptogenesis. For $n^N_{\alpha\beta} = 0$, the RH neutrino masses are comparable with one another and given by
\begin{align}
	m_{N_\alpha}= 
	y^N_\alpha M
	\simeq 8.7\times 10^{13}\,\text{GeV}~y^N_\alpha
	\left(\frac{1}{\kappa\Delta} \right) 
	\left(\frac{\mathcal F/\Lambda}{0.8}\right)
	\left(\frac{50}{N_e}\right).
\end{align}
These are a little bit larger than $T_R$ in \eqref{reheating_temperature}, which reduces the final baryon asymmetry in general. However, it is known that in the strong washout regime the suppression is not significant as long as $T_R\gtrsim \mathcal O(0.1)m_{N_\alpha}$~\cite{Buchmuller:2004nz}. Noting a relatively large $m_{N_\alpha}$, one can see from \eqref{baryon_asymmetry_susy} that a sufficient amount of baryon asymmetry can be obtained for $\kappa\Delta\gtrsim 1$.\footnote{In the present charge assignment, the RH neutrino masses can be degenerate, which enhances the baryon asymmetry~\cite{Covi:1996fm,Pilaftsis:1997dr,Buchmuller:1997yu,Pilaftsis:2003gt,Garny:2011hg}.
}
Thanks to the lepton parity and the charge assignment mentioned in Sec.~\ref{sec:gravitino}, the high reheating temperature does not cause the gravitino/axino problem.

To summarize, we can realize a consistent cosmological scenario after inflation in our inflation model. Although the existence of the sflaxion constrains model parameters, reheating, thermal leptogenesis, and the flaxion dark matter can be viable by taking $n^N_{\alpha\beta} = 0$ and $\Lambda\sim \mathcal{O}(10^{13})~\text{GeV}$.

\section{Summary and discussion}
\label{sec:summary}

In this paper, we have studied a SUSY extension of the flaxion model
in which the NG boson in association with the spontaneous breaking of
the U(1)$_\text{F}$ flavor symmetry plays the role of the axion.  As
in the case of the non-SUSY flaxion model, we have shown that (i) the
Yukawa structures in the quark and lepton sectors, as well as the
active neutrino mass matrix, can be well explained with a proper
choice of the U(1)$_\text{F}$ charges of the fields in the MSSM as
well as those of right-handed (s)neutrinos, (ii) coherent oscillation
of the flaxion can be dark matter, and (iii) the flavor violating
decay of $K^+\rightarrow\pi^+ a$ is an interesting check point.  We
have also discussed CP and flavor violations due to SUSY particles and
the effects of $R$-parity violation, which are new potential problems
due to the SUSY extension.
We have also shown that the cosmological gravitino and axino problems can be avoided by a small $R$-parity violation, which is also controlled by the U(1)$_\text{F}$ charges.

Compared to the non-SUSY case, one of the new important issues in the SUSY
flaxion model is the light sflaxion; in the SUSY model, the mass of
the sflaxion, which is the real component of the complex scalar field
containing the flaxion, is expected to be of the order of the
gravitino mass which is much smaller than the breaking scale of the
flavor symmetry.  The cosmological evolution of such a light field is
non-trivial, and may spoil the standard thermal history.  
We have studied the evolution of the sflaxion field in the early universe.
In particular, because of thermal effects, the potential of the sflaxion
may be significantly deformed in the early universe, which may result
in the trapping of the sflaxion field in a minimum of the potential
much far away from the sflaxion amplitude in the vacuum (which
corresponds to the absolute minimum of the potential at zero
temperature).  We have shown that there exists a viable parameter
space in which the thermal effects are insignificant.

We have also pointed out that the flaxion multiplets can play the role
of inflaton in an attractor-like inflation scenario.  Adopting a particular
form of the K\"ahler potential, we have studied the dynamics of the
inflaton and other scalar fields during and after inflation taking
into account the effects of supergravity, and have shown that a viable cosmological
scenario is possible, including inflation, leptogenesis, and the dark matter.

\section*{Acknowledgement}
This work was supported in part by the Grant-in-Aid for Scientific Research A (26247038 [KH], 26247042 [KN], 16H02189 [KH]), Scientific Research C (26400239 [TM]), Young Scientists B (26800121 [KN], 26800123 [KH]) and Innovative Areas (26104001 [KH], 26104009 [KH and KN], 15H05888 [KN], 17H06359 [KN], 16H06490 [TM]), 
by JSPS Research Fellowships for Young Scientists [YE],
by the Program for Leading Graduate Schools [YE], 
and by World Premier International Research Center Initiative (WPI Initiative), MEXT, Japan.

\bibliographystyle{JHEP}
\bibliography{ref}

\end{document}